\documentclass[twocolumn,english,aps,floats]{revtex4}
\usepackage[T1]{fontenc}
\usepackage[latin1]{inputenc}
\usepackage{babel}
\usepackage{graphics}

\makeatletter

\providecommand{\LyX}{L\kern-.1667em\lower.25em\hbox{Y}\kern-.125emX\@}
\let\SF@@footnote\footnote
\def\footnote{\ifx\protect\@typeset@protect
    \expandafter\SF@@footnote
  \else
    \expandafter\SF@gobble@opt
  \fi
}
\expandafter\def\csname SF@gobble@opt \endcsname{\@ifnextchar[
  \SF@gobble@twobracket
  \@gobble
}
\edef\SF@gobble@opt{\noexpand\protect
  \expandafter\noexpand\csname SF@gobble@opt \endcsname}
\def\SF@gobble@twobracket[#1]#2{}

\usepackage[T1]{fontenc}
\usepackage[latin1]{inputenc}
\usepackage{graphics}

\usepackage{amsmath}
\usepackage{amssymb}

\makeatletter

\usepackage{graphics}

\makeatother

\makeatother
\begin{document}

\title{Stimulated Raman adiabatic passage from an atomic to a molecular
Bose-Einstein condensate}

\author{P. D. Drummond and K. V. Kheruntsyan}

\affiliation{\textit{Department of Physics, The University of Queensland, Brisbane,}\\
 Queensland 4072, Australia}

\author{D. J. Heinzen and R. H. Wynar}

\affiliation{\textit{Department of Physics, University of Texas, Austin, Texas,
78712}}

\date{\today {}}

\begin{abstract}
The process of stimulated Raman adiabatic passage (STIRAP) provides
a possible route for the generation of a coherent molecular Bose-Einstein
condensate (BEC) from an atomic BEC. We analyze this process in a
three-dimensional mean-field theory, including atom-atom interactions
and non-resonant intermediate levels. We find that the process is
feasible, but at larger Rabi frequencies than anticipated from a crude
single-mode lossless analysis, due to two-photon dephasing caused
by the atomic interactions. We then identify optimal strategies in
STIRAP allowing one to maintain high conversion efficiencies with
smaller Rabi frequencies and under experimentally less demanding conditions.


\end{abstract}
\maketitle

\section{Introduction}

Coherent conversion of an atomic to a molecular Bose-Einstein condensate
(BEC) is a first step towards `superchemistry' \cite{HWDK2000}, which
is the stimulated emission of molecules in a chemical reaction. A
number of studies of this \cite{KD98,DKH98,JM99} have shown that
direct conversion via Raman photoassociation \cite{WFHRH2000} appears
feasible, based on stimulated free-bound and bound-bound transitions
in the presence of two laser fields of different frequencies \cite{Raman-photoassociation}.
Here pairs of atoms from the two-atom continuum of the ground electronic
potential are transferred -- via an excited bound molecular state
-- to a bound molecular state of a lower energy in the ground potential.
Raman photoassociation allows coupling to a single molecular state,
which can be selected by the Raman laser frequencies. Practical estimates
using available lasers and transitions indicate that coherent transfer
may be limited by spontaneous emission from the intermediate molecular
excited electronic state. Another mechanism that can result in coupled
atomic-molecular BEC systems \cite{AMBEC-Feshbach} is based on Feshbach
resonances \cite{Feshbach}. However, realistic analysis and experimental
implementations \cite{Feshbach-experiment} indicate that the loss
processes due to inelastic atom-molecule collisions occur at a significant
rate.

A possible route towards minimizing losses and decoherence from spontaneous
emission in photoassociation is stimulated Raman adiabatic passage
(STIRAP) \cite{MKJ2000}, in which a counter-intuitive pulse sequence
is used, where the first input pulse couples the molecular levels
- even when there are no molecules present. In this situation, a dark
superposition state is formed, due to interference effects between
the atomic and molecular electronic ground states. This minimizes
the probability of a real transition to the molecular excited state,
and hence reduces spontaneous emission. Previous analyses of this
problem have not taken into account losses, collisions, or the full
three-dimensional structure of the two Bose condensates in a trap.

In this paper, we provide an analysis which is relevant in a physically
appropriate model that does include the known physics of spontaneous
emission losses, \( s \)-wave scattering processes and spatial diffusion
of the condensates. The result is that the STIRAP process appears
feasible at high laser pulse intensities, provided the Rabi frequency
is much greater than the two-photon detuning due to mean-field interactions.
We give a detailed numerical calculation based on both homogeneous
and inhomogeneous three-dimensional mean-field (Gross-Pitaevskii type)
theories, including couplings to non-resonant intermediate levels,
and show how the results scale with the two-photon detuning, pulse
duration and Raman pulse intensities. An optimal situation is found
by considering an off-resonance operation and different effective
Rabi frequencies in the two Raman channels. We show that these strategies
can greatly enhance the conversion efficiency for given laser intensities,
thus making the experimental requirements much more feasible.

\section{Coupled Gross-Pitaevskii equations for STIRAP}

We start by considering the theory of coherently interacting atomic
and molecular condensates needed to describe this process \cite{HWDK2000,DKH98},
and assume a specific coupling mechanism based on stimulated free-bound
Raman transitions \cite{Raman-photoassociation}, in which two atoms
of energy \( E_{1} \) collide to from a molecule of energy \( E_{2} \)
, with an excited molecular state forming as an intermediate step.
The Raman coupling is induced by two laser fields of frequencies \( \omega _{1} \)
and \( \omega _{2} \), and becomes resonant when the Raman detuning
\( \delta =(2E_{1}-E_{2})/\hbar -(\omega _{2}-\omega _{1}) \) goes
to zero. This allows coupling to a single molecular state, which can
be selected by the Raman laser frequencies.

We derive the atom-molecule coupling for a simplified model of the
two-body interaction \cite{HWDK2000,photoassociation}, in which the
atoms interact in their electronic ground state through a single Born-Oppenheimer
potential \( V_{g}(R) \). Molecules are formed in a single bound
vibrational state of energy \( E_{2} \) with radial wave function
\( u_{2}(R) \). Two free atoms with zero relative kinetic energy
have a total energy \( 2E_{1} \), and a relative radial wave function
\( u_{1}(R) \), normalized so that asymptotically \( u_{1}\propto (1-a_{1}/R) \).
We assume that the laser field has two frequency components, with
\( {\mathbf{E}}=\mathop {\textrm{Re}}\sum \left[ {\mathbf{E}}^{(i)}\exp (i\omega _{i}t)\right]  \),
\( i=1,2 \). Each couples the ground electronic state to a single
electronically excited state described by a potential \( V_{e}(R) \),
with `bare' electronic Rabi frequencies \( \Omega _{i}^{(el)}({\mathbf{R}})={\mathbf{d}}_{3i}({\mathbf{R}})\cdot {\mathbf{E}}^{(i)}/\hbar  \)
, where \( {\mathbf{d}}_{3i}({\mathbf{R}}) \) is the molecular electric
dipole matrix element with a nuclear separation of \( {\mathbf{R}} \).
The excited state has vibrational levels \( |v^{\prime }\rangle  \)
with energies \( E_{v^{\prime }} \) and radial wave functions \( u_{v^{\prime }}(R) \).
All bound levels are normalized so that \( \int d^{3}{\mathbf{R}}|u_{v^{\prime }}|^{2}=\int d^{3}{\mathbf{R}}|u_{2}|^{2}=1 \).

\begin{figure}
\par\centering \resizebox*{6cm}{!}{\includegraphics{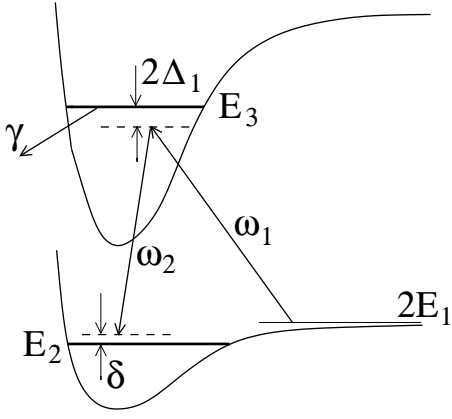} } \par{}
\caption{Diagrammatic representation of the free-bound and bound-bound transitions
in STIRAP.}
\label{Diagram}
\end{figure}

We consider the case of near resonant transitions \( 1\rightarrow v^{\prime }\rightarrow 2 \)
and denote the resonant excited vibrational level \( v^{\prime } \)
via index \( 3 \) (see Fig. \ref{Diagram}). The usual quantum field
theory Hamiltonian \cite{Abrikosov-Dzyaloshinski} for noninteracting
atomic (\( i=1 \)) and molecular (\( i=2,3 \)) species, in well-defined
internal states described by annihilation operators \( \hat{\Psi }_{i} \),
is given by: \begin{equation}
\hat{H}^{(0)}=\int d^{3}{\mathbf{x}}\sum _{i=1}^{3}\left[ \frac{\hbar ^{2}}{2m_{i}}|\nabla \hat{\Psi }_{i}({\mathbf{x}})|^{2}+V_{i}({\mathbf{x}})\hat{\Psi }_{i}^{\dag }({\mathbf{x}})\hat{\Psi }_{i}^{{}}({\mathbf{x}})\right] .
\end{equation}
 Here \( m_{i} \) (\( i=1,2,3 \)) are the masses, \( m_{2,3}=2m_{1} \),
and \( V_{i}({\mathbf{x}}) \) is the trapping potential including
the internal energy for the \( i \)-th species, where we define \( V_{i}(0)=E_{i} \).

Including \( s \)-wave scattering processes and laser induced particle
inter-conversion, together with the assumption of a momentum cut-off,
results in the following terms in the effective interaction Hamiltonian:
\begin{eqnarray}
\hat{H}_{int}^{(s)} & = & \frac{\hbar }{2}\int d^{3}{\mathbf{x}}\sum _{ij}U_{ij}\hat{\Psi }_{i}^{\dag }({\mathbf{x}})\hat{\Psi }_{j}^{\dagger }({\mathbf{x}})\hat{\Psi }_{j}^{{}}({\mathbf{x}})\hat{\Psi }_{i}^{{}}({\mathbf{x}}),\\
\hat{H}_{int}^{(1-3)} & = & \int d^{3}{\mathbf{x}}\left[ \frac{-\hbar {\Omega }_{1}}{2\sqrt{2}}e^{-i\omega _{1}t}\hat{\Psi }_{1}^{{2}}({\mathbf{x}})\hat{\Psi }_{3}^{\dagger }({\mathbf{x}})+H.c.\right] ,\\
\hat{H}_{int}^{(2-3)} & = & \int d^{3}{\mathbf{x}}\left[ \frac{-\hbar {\Omega }_{2}}{2}e^{-i\omega _{2}t}\hat{\Psi }_{2}^{{}}({\mathbf{x}})\hat{\Psi }_{3}^{\dagger }({\mathbf{x}})+H.c.\right] .
\end{eqnarray}
 Here \( \Omega _{i}=\int d^{3}{\mathbf{R}}\Omega _{i}^{(el)}({\mathbf{R}})\, u_{3}^{\ast }(R)u_{i}(R)\approx \overline{\Omega }_{i}^{(el)}I_{i,3} \)
(\( i=1,2 \))\ are the molecular Rabi frequencies. These can be treated
using the Franck-Condon overlap integrals \( I_{i,3}=\int d^{3}{\mathbf{R}}u_{3}^{\ast }(R)u_{i}(R) \),
if we take \( \overline{\Omega }_{i}^{(el)} \) as the mean electronic
Rabi frequency. We note that \( \Omega _{1} \), which connects the
atomic and molecular condensates, has units of \( s^{-1}m^{-3/2} \),
and must be multiplied by the atomic condensate amplitude to obtain
a true Rabi frequency. The couplings \( U_{ij} \) in the diagonal
case are given by \( U_{ii}=4\pi \hbar a_{i}/m_{i} \), where \( a_{i} \)
is the \( i \)-th species \( s \)-wave scattering length, while
the non-diagonal terms are given by \( U_{ij}=U_{ji}=2\pi \hbar a_{ij}/\mu _{ij} \),
where \( a_{ij} \) is the inter-species scattering length and \( \mu =m_{i}m_{j}/(m_{i}+m_{j}) \)
is the reduced mass.

In addition, we account for losses from each state, at a rate \( \gamma _{i} \).
The resulting Heisenberg equations for the field operators are treated
within the mean-field approximation, in which the operators are replaced
by their mean values, and a factorization is assumed. This approximation
is expected to be valid at sufficiently high density. Corrections
due to quantum correlations \cite{Holland} have been treated in greater
detail elsewhere \cite{Hope-Olsen-Plimak}. Next, we introduce rotating
frame detunings, defined so that: \begin{eqnarray}
2\Delta _{1}({\mathbf{x}}) & = & (E_{3}-2V_{1}({\mathbf{x}}))/\hbar -\omega _{1}\, \, ,\nonumber \\
\Delta _{2}({\mathbf{x}}) & = & (E_{3}-V_{2}({\mathbf{x}}))/\hbar -\omega _{2}\, \, ,\\
\Delta _{3}({\mathbf{x}}) & = & (E_{3}-V_{3}({\mathbf{x}}))/\hbar \, \, .\nonumber 
\end{eqnarray}
 In the case of uniform condenstates, \( V_{i}({\mathbf{x}}) \) are
equal to \( V_{i}({\mathbf{x}})=E_{i} \), and \( \Delta _{3}=0 \).

This results in the following set of Gross-Pitaevskii type of equations
for the mean-field amplitudes, in rotating frames such that \( \psi _{i}=\left\langle \hat{\Psi }_{i}\right\rangle \exp [i(E_{i}+\Delta _{i}(0))t/\hbar ] \):
\begin{eqnarray}
\frac{\partial \psi _{1}({\mathbf{x}},t)}{\partial t} & = & i\Delta _{1}^{GP}\psi _{1}+\frac{i{\Omega }^{\ast }_{1}}{\sqrt{2}}\psi _{3}\psi _{1}^{\ast },\nonumber \\
\frac{\partial \psi _{2}({\mathbf{x}},t)}{\partial t} & = & i\Delta _{2}^{GP}\psi _{2}+\frac{i{\Omega }^{\ast }_{2}}{2}\psi _{3},\label{psi_{3}} \\
\frac{\partial \psi _{3}({\mathbf{x}},t)}{\partial t} & = & i\Delta _{3}^{GP}\psi _{3}+\frac{i{\Omega }_{1}}{2\sqrt{2}}\psi _{1}^{2}+\frac{i{\Omega }_{2}}{2}\psi _{2},\nonumber 
\end{eqnarray}
 Here \( \Delta _{j}^{GP} \) is the \( i \) -th Gross-Pitaevskii
mean-field detuning in the rotating frame, defined so that: \begin{equation}
\label{GP-detuning}
\Delta _{j}^{GP}({\mathbf{x}},t)=\Delta _{j}({\mathbf{x}})+\frac{\hbar }{2m_{j}}\nabla ^{2}-\sum _{k=1}^{3}U_{jk}|\psi _{k}|^{2}+i\frac{\gamma _{j}}{2}\, \, .
\end{equation}
 We also introduce the two-photon laser detuning at trap center: \begin{equation}
\label{Two-photon-detuning}
\delta \equiv \Delta _{2}(0)-2\Delta _{1}(0)=-(E_{2}-2E_{1})/\hbar +(\omega _{1}-\omega _{2}).
\end{equation}

In addition to losses due to spontaneous emission from the electronic
excited states, rotationally or vibrationally inelastic atom-molecule
collisions may also give rise to losses. The magnitude of these rates
is presently unknown, and we neglect them here. We note that these
rates should decrease rapidly with increasing molecular binding energy
and go to zero in the molecular ground state, so that it should be
possible to obtain a very low rate by selecting the coupling to a
deeply bound molecular level. This approximation means that we will
set \( \gamma _{i}=\gamma \delta _{3i} \) in the following treatment,
where \( \delta _{ij} \) is the Kronecker delta-function.

The simplest STIRAP scheme employs exact tuning of the laser frequencies
to the bare state resonances, i.e., \( \Delta _{1}(0)=\Delta _{2}(0)=0 \).
This, however, is not necessarily required as the STIRAP can still
occur with detuned intermediate levels. Moreover, as we show below,
in BEC environments, the off-resonance operation turns out to be more
efficient if the detunings compensate for the phase shifts due to
mean-field energies. In effect this is equivalent to a renormalized
two-photon {}``on-resonant'' operation in which \( \Delta ^{GP}_{2}({\mathbf{x}},0)-2\Delta ^{GP}_{1}({\mathbf{x}},0)\approx 0 \).

If, instead of considering near-resonance coupling, we consider a
large intermediate level detuning so that the excited state can be
adiabatically eliminated, we recover the basic terms in the set of
equations analyzed in \cite{HWDK2000}. Simultaneously, this would
give us the previously known result \cite{photoassociation} for the
laser induced modification to the scattering length \( a_{1} \) occurring
in a simple single-laser photoassociation of pairs of atoms, thus
justifying the above form of the interaction Hamiltonian. In the present
paper, however, we assume that the detunings \( \Delta _{i} \) are
small compared to the characteristic separation between the vibrational
levels so that all other vibrational levels can be neglected. The
near-resonant excited level is treated explicitly, rather than eliminated
adiabatically as in \cite{HWDK2000}.

\section{STIRAP in a BEC}

Before carrying out simulations of the full \( 3D \) equations in
a trap, it is instructive to start with a simplified version of the
theory - in which there are no kinetic energy terms. We expect this
approximation to be valid in the Thomas-Fermi limit of large, relatively
dense condensates, which is a regime of much current experimental
interest. This is described by the following set of equations: \begin{eqnarray}
\frac{\partial \psi _{1}({\mathbf{x}},t)}{\partial t} & = & i\Delta _{1}^{TF}\psi _{1}+\frac{i{\Omega }^{\ast }_{1}}{\sqrt{2}}\psi _{3}\psi _{1}^{\ast },\nonumber \\
\frac{\partial \psi _{2}({\mathbf{x}},t)}{\partial t} & = & i\Delta _{2}^{TF}\psi _{2}+\frac{i{\Omega }^{\ast }_{2}}{2}\psi _{3},\label{TF} \\
\frac{\partial \psi _{3}({\mathbf{x}},t)}{\partial t} & = & i\Delta _{3}^{TF}\psi _{3}+\frac{i{\Omega }_{1}}{2\sqrt{2}}\psi _{1}^{2}+\frac{i{\Omega }_{2}}{2}\psi _{2},\nonumber 
\end{eqnarray}
 where we have introduced as an effective Thomas-Fermi limit frequency
shift, \begin{equation}
\label{Thomas-Fermi}
\Delta _{j}^{TF}({\mathbf{x}},t)=\Delta _{j}({\mathbf{x}})-\sum _{k=1}^{3}U_{jk}|\psi _{k}|^{2}+i\frac{\gamma _{j}}{2}\, \, .
\end{equation}
 This corresponds to the treatment given in \cite{MKJ2000}, except
that we explicitly include the loss term \( \gamma  \) due to spontaneous
emission, the \( s \)-wave scattering processes due to \( U_{ij} \),
and the Franck-Condon integrals into the coupling coefficients for
the free-bound and bound-bound transitions.

In order to understand how this is related to the usual STIRAP technique
in a three-level \( \Lambda  \) atomic system, we introduce a new
wave-function \( \psi _{m}=\psi _{1}/\sqrt{2} \), which corresponds
to the coherent amplitude of a (virtual) molecular condensate with
the same number of atoms as in the atomic BEC. We then introduce a
Bose-stimulated Rabi frequency, which includes a local coherent BEC
amplitude for the first free-bound transition: \( \widetilde{\Omega }_{1}=\psi _{1}^{\ast }{\Omega }_{1} \).
This leads to the equations: \begin{eqnarray}
\frac{\partial \psi _{m}({\mathbf{x}},t)}{\partial t} & = & i\Delta _{1}^{TF}\psi _{m}+\frac{i\widetilde{\Omega }^{\ast }_{1}}{2}\psi _{3},\nonumber \\
\frac{\partial \psi _{2}({\mathbf{x}},t)}{\partial t} & = & i\Delta _{2}^{TF}\psi _{2}+\frac{i\Omega ^{\ast }_{2}}{2}\psi _{3},\\
\frac{\partial \psi _{3}({\mathbf{x}},t)}{\partial t} & = & i\Delta _{3}^{TF}\psi _{3}+\frac{i\widetilde{\Omega }_{1}}{2}\psi _{m}+\frac{i\Omega _{2}}{2}\psi _{2}\, \, .\nonumber 
\end{eqnarray}
 These are precisely the usual STIRAP equations, except with additional
detunings coming from the mean-field interactions, and a Rabi frequency
in the first transition which is proportional to the amplitude of
the atomic BEC wave-function. In practise, the Rabi frequency may
have an additional space-dependence due to the spatial variation of
the laser phase and intensity. We therefore conclude that, provided
we can satisfy the normal adiabatic STIRAP requirements of slow time-variation
in the \emph{effective} Rabi frequencies in the above equations, the
technique will also work for a BEC. This is a simpler proof than previously
\cite{MKJ2000}. In particular, we can immediately deduce the expected
solution for real Rabi frequencies in the adiabatic limit. In order
to have \( \widetilde{\Omega }_{1}\psi _{m}+\Omega _{2}\psi _{2}=0 \),
one requires: \begin{eqnarray}
\psi _{1}({\mathbf{x}},t) & = & \psi _{1}({\mathbf{x}},0)\cos (\theta ),\nonumber \\
\psi _{2}({\mathbf{x}},t) & = & -\psi _{1}({\mathbf{x}},0)\sin (\theta )/\sqrt{2},\\
\psi _{3}({\mathbf{x}},t) & = & 0.\nonumber 
\end{eqnarray}

Here the space dependent mixing angle \( \theta ({\mathbf{x}},t) \)
is obtained from the ratio of effective Rabi frequencies: \begin{equation}
\label{solution}
\tan (\theta )=\frac{\widetilde{\Omega }_{1}}{\Omega _{2}}=\frac{\psi _{1}\Omega _{1}}{\Omega _{2}}=\left[ \sqrt{\frac{\psi ^{2}_{1}({\mathbf{x}},0)\Omega _{1}^{2}}{\Omega ^{2}_{2}}+\frac{1}{4}}-\frac{1}{2}\right] ^{1/2}\, .
\end{equation}

We can see that \emph{initially}, while \( \psi _{1} \) is still
close to its initial value, the mixing angle is close to its expected
value in normal STIRAP, since \( \tan (\theta )\approx \psi _{1}({\mathbf{x}},0)\Omega _{1}/\Omega _{2} \).
However, at the final stages of the adiabatic passage, the nonlinear
effects due to the atom-molecular coupling become important. As the
atomic BEC amplitude only varies on the time-scale of the input fields,
the nonlinear atom-molecular coupling term by itself should not introduce
new adiabatic restrictions. However, there will be time-dependent
detunings introduced by the mean-field terms.

Finally, we can see that similar conclusions can also be reached in
a non-uniform BEC, by replacing the uniform detunings with appropriate
Gross-Pitaevskii detunings, that include the spatial potentials. In
a typical BEC cooled in the Thomas-Fermi regime, we expect the kinetic
energy terms to have relatively small effects.

STIRAP can therefore be implemented as usually by using two laser
pulses applied in counterintuitive order. We choose Gaussian pulses
of the form \begin{equation}
\bar{\Omega }_{i}^{(el)}(t)=\Omega _{i}^{(el,0)}\exp [-(t-t_{i})^{2}/T^{2}],\; \; \; \; (i=1,2),
\end{equation}
 or \begin{equation}
\Omega _{i}(t)=\Omega _{i}^{(0)}\exp [-(t-t_{i})^{2}/T^{2}],
\end{equation}
 where the peak values are related as follows: \( \Omega _{i}^{(0)}=\Omega _{i}^{(el,0)}I_{i,3} \).
The pulse at frequency \( \omega _{2} \) is applied first, with the
center at \( t_{2} \), while the second pulse at frequency \( \omega _{1} \)
is delayed by \( \alpha T \), i. e., \begin{equation}
t_{1}-t_{2}=\alpha T,
\end{equation}
 where \( \alpha  \) is the delay coefficient, and \( T \) is the
pulse duration, which we assume is the same for both pulses.

In terms of the Rabi frequencies, the adiabatic condition for STIRAP
now reads as \cite{Bergmann}: \begin{equation}
\label{adiabatic-1}
\Omega (t)\Delta \tau \gg \sqrt{1+|2\Delta _{1}^{TF}-\Delta _{3}^{TF}|\Delta \tau },
\end{equation}
 where \( \Omega (t)=\sqrt{|\widetilde{\Omega }_{1}(t)|^{2}+|\Omega _{2}(t)|^{2}} \)
is the rms Rabi frequency, \( \Delta \tau  \) is the duration during
which the pulses overlap, and \( (2\Delta _{1}^{TF}-\Delta _{3}^{TF}) \)
simply corresponds to the detuning of the single-photon transition.
However, there is a second condition, which is often not stated explicitly.
This is that STIRAP requires an effective two-photon resonance, to
avoid dephasing between the initial and final states in the dark-state
superposition. The two-photon resonance condition is different from
the usual STIRAP case, since a detuning of \( \Delta _{1}^{TF} \)
causes a phase rotation \emph{both} in \emph{\( \psi _{m} \)} and
in \emph{\( \widetilde{\Omega }_{1} \)} as well, since this also
includes a phase term from the condensate. As a result, the necessary
condition for two-photon resonance is therefore: \begin{equation}
\label{adiabatic-2}
|\Delta _{2}^{TF}-2\Delta _{1}^{TF}|\Delta \tau \ll 1\, \, .
\end{equation}
 This leads to a third condition, which shows that there is a lower
bound to the allowed Rabi frequency in order to have STIRAP occurring
in the presence of mean-field dephasing effects: \begin{equation}
\label{adiabatic-3}
\Omega (t)\gg |\Delta _{2}^{TF}-2\Delta _{1}^{TF}|\, \, .
\end{equation}

As is usually the case in STIRAP, these conditions cannot be satisfied
very early or late in the pulse sequence, when the Rabi frequencies
are small; but they should be satisfied over most of the STIRAP interaction,
and over most of the condensate volume. For simplicity, we will apply
these conditions to the peak Rabi frequency \( \Omega ^{(0)} \),
and to the total pulse duration \( T \). Further, since \( \widetilde{\Omega }_{1} \)
is itself a function of the STIRAP evolution, we introduce an effective
first Rabi frequency, defined in terms of the initial density \( n_{1}(0)=|\psi _{1}(0)|^{2} \).
This is an upper bound to the stimulated Rabi frequency; thus \( \Omega _{1}^{(eff)}(t)=\sqrt{n_{1}(0)}\Omega _{1}(t)\geq \widetilde{\Omega }_{1}(t) \)
(and sometimes we write \( \Omega _{2}^{(eff)}(t)=\Omega _{2}(t) \),
for uniformity), where \begin{equation}
\label{pulse-shape}
\Omega _{i}^{(eff)}(t)=\Omega _{i}^{(eff,0)}\exp [-(t-t_{i})^{2}/T^{2}],\; \; \; \; (i=1,2)\, ,
\end{equation}
 where \( \Omega _{1}^{(eff,0)}=\Omega _{1}^{(el,0)}I_{1,3}\sqrt{n_{1}(0)} \)
and \( \Omega _{2}^{(eff,0)}=\Omega _{2}^{(el,0)}I_{2,3} \).

\begin{figure}
\par\centering \resizebox*{6cm}{!}{\includegraphics{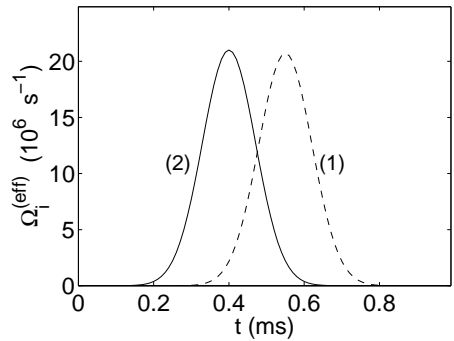} } \par{}
\caption{The Rabi frequencies \protect\( \Omega ^{(eff)}_{i}(t)\protect \)
for the Raman transitions, with the peak values of \protect\( \Omega ^{(eff,0)}_{1}=\Omega ^{(eff,0)}_{2}=2.1\times 10^{7}\protect \)
s\protect\( ^{-1}\protect \), pulse durations \protect\( T=10^{-4}\protect \)
s, and a delay coefficient of \protect\( \protect \alpha =1.5\protect \).}
\label{STIRAP-pulses}
\end{figure}

A typical pulse sequence is shown in Fig. \ref{STIRAP-pulses}. From
the definition of \( \Omega _{i} \), the Franck-Condon overlap integrals
are an important issue from the point of view of employing a realistic
set of parameters. Since the overlap integrals enter into the definition
of the effective Rabi frequencies, their values will affect the adiabatic
condition rewritten in terms of the {}``bare'' electronic Rabi frequencies
\( \overline{\Omega }_{i}^{(el)} \). We will analyze this in more
detail in the next section.

\section{Uniform condensate results}

We start by considering a uniform condensate, described by a similar
equation to the Thomas-Fermi case, except with a uniform trap potential
for simplicity:

\begin{eqnarray}
\frac{\partial \psi _{m}({\mathbf{x}},t)}{\partial t} & = & i\Delta _{1}^{TF}(0)\psi _{m}+\frac{i\widetilde{\Omega }^{\ast }_{1}}{2}\psi _{3},\nonumber \\
\frac{\partial \psi _{2}({\mathbf{x}},t)}{\partial t} & = & i\Delta _{2}^{TF}(0)\psi _{2}+\frac{i\Omega ^{\ast }_{2}}{2}\psi _{3},\label{J-3} \\
\frac{\partial \psi _{3}({\mathbf{x}},t)}{\partial t} & = & i\Delta _{3}^{TF}(0)\psi _{3}-\frac{i\widetilde{\Omega }_{1}}{2}\psi _{m}-\frac{i\Omega _{2}}{2}\psi _{2}\, \, .\nonumber 
\end{eqnarray}

Here the uniform detuning term \( \Delta _{1}^{TF}(0) \) is defined
as the Thomas-Fermi detuning, evaluated at the trap center. We start
by considering a uniform condensate in which the \( s \)-wave scattering
interactions are negligible (i.e., \( U_{ij}=0 \)), and assume exact
resonances with respect to bare state transitions, \( \Delta _{1}=\Delta _{2}=0 \).
We first simulate the above simplified equations (\ref{J-3}) with
an initial condition of a pure atomic condensate. This model is more
realistic than that of M. Mackie et al. \cite{MKJ2000}, as it includes
spontaneous emission. We find that including the loss term \( \gamma  \)
imposes restrictions on the effective Rabi frequencies \( \Omega _{i}^{(eff,0)} \)
and the duration of the pulses \( T \).

The results are best understood with reference to Table \ref{STIRAP_param},
which gives the values of typical STIRAP parameters characteristic
of a condensate of \( ^{87} \)Rb atoms \cite{HWDK2000,WFHRH2000}),
and corresponding to the pulse sequence in Fig. \ref{STIRAP-pulses}. 
\begin{table}
\begin{tabular}{|l|l|}
\hline 
\( \gamma  \)&
 \( 7.4\times 10^{7} \) s\( ^{-1} \)\\
\hline
\( n_{1}(0) \)&
 \( 4.3\times 10^{20} \) m\( ^{-3} \)\\
\hline
\( \Omega _{1}^{(eff,0)}=\Omega _{2}^{(eff,0)} \)&
 \( 2.1\times 10^{7} \) s\( ^{-1} \)\\
\hline
\( T \)&
 \( 10^{-4} \) s  \\
\hline
\end{tabular}

\caption{Typical parameter values for efficient STIRAP.}

\label{STIRAP_param}
\end{table}

Taking the values of the parameters in Table \ref{STIRAP_param},
and an optimum delay coefficient of \( \alpha \simeq 1.5 \), gives
\( \eta \simeq 0.96 \) or about \( 96\% \) efficiency of conversion
of atomic BEC into molecular BEC, even including the upper level spontaneous
emission. Here, the conversion efficiency \( \eta  \) is defined
as the fraction of the initial number of atoms \( n_{1}(0) \) converted
into molecules \begin{equation}
\label{efficiency}
\eta =\frac{2n_{2}(\infty )}{n_{1}(0)}\, \, ,
\end{equation}
 where \( n_{2}(\infty ) \) is the final number of molecules. This
accounts for the fact that \( n_{1} \) atoms can produce \( n_{1}/2 \)
molecules at best.

For comparison, using smaller Rabi frequencies, \( \Omega _{i}^{(eff,0)}=2.1\times 10^{6} \)
s\( ^{-1} \), and a larger value of \( T=10^{-3} \) s, so that the
product \( \Omega _{i}^{(eff,0)}T \) still has the previous value
\( \Omega _{i}^{(eff,0)}T=2.1\times 10^{3} \), gives a maximum conversion
efficiency of \( \eta \simeq 0.83 \), with a new optimum delay coefficient
\( \alpha \simeq 1.2 \). This is smaller than in the previous example.
In order to reach the same efficiency as before, one has to further
increase the pulse durations (up to \( T=10^{-2} \) s), i.e. enter
into a deeper adiabatic regime. In in the absence of the spontaneous
emmission term, the conversion efficiencies would not be different
in these two examples.

In other words, in this simplified model it is possible to have effective
Rabi frequencies smaller than the spontaneous emission rate \( \gamma  \),
provided the duration of the pulses is long enough. As usual, we can
understand this physically as implying that the upper level is never
actually occupied for very slowly varying adiabatic pulses. Hence,
just as in the case of atomic STIRAP, we can ignore spontaneous emission
from the upper level provided that we use very slowly varying pulses
which are sufficiently deep in the adiabatic limit. As we see in the
following calculations, the problem with this strategy is that very
long pulses will tend to cause violations of the two-photon resonance
condition, in the presence of mean-field interactions.

\subsection{Effects of the mean-field energies}

We now wish to include the mean-field energy terms, and first restrict
our analysis to the atom-atom scattering processes. We consider a
characteristic value of \( U_{11}=4.96\times 10^{-17} \) m\( ^{3} \)/s
corresponding to the scattering length of \( ^{87} \)Rb atoms \cite{Heinzen-Rb87-scatteringlength}
\( a_{1}=5.4 \) nm (\( m_{1}=1.443\times 10^{-25} \) kg). Together
with the choice of the initial atomic density \( n_{1}(0) \) as before
(see Table \ref{STIRAP_param}), the value of \( U_{11} \) sets up
a characteristic dephasing time scale \begin{equation}
\label{t_{h}}
t_{ph}=\left[ U_{11}n_{1}(0)\right] ^{-1},
\end{equation}
 equal in this case to \( t_{ph}\simeq 4.7\times 10^{-5} \) s. The
pulse duration \( T \) must be smaller than or of the order of the
dephasing time, in order to permit STIRAP, otherwise the two-photon
resonance condition will not be satisfied.

Thus, including atom-atom scattering imposes an upper limit to the
pulse duration, so that \( T \) has now to satisfy \( T\lesssim t_{ph}\simeq 10^{-5}-10^{-4} \)
s. But this restriction means that one can no longer use \emph{longer}
pulse durations for \emph{small} values of \( \Omega ^{(eff,0)}_{i} \),
while still maintaining high conversion efficiency. As a result, the
adiabatic condition \( \Omega ^{(eff,0)}_{i}T\gg 1 \), with the restriction
of \( T\lesssim 10^{-5}-10^{-4} \) s, requires high peak values of
the effective Rabi frequencies: \( \Omega ^{(eff,0)}_{i}\gtrsim 10^{7} \)
s\( ^{-1} \).

\begin{table}
\begin{tabular}{|l|l|}
\hline 
\( U_{11} \)&
 \( 4.96\times 10^{-17} \) m\( ^{3} \)/s \\
\hline
\( U_{12} \)&
 \( -6.44\times 10^{-17} \) m\( ^{3} \)/s \\
\hline
\( U_{22} \)&
 \( 2.48\times 10^{-17} \) m\( ^{3} \)/s \\
\hline
\( U_{3i} \)&
 \( 0 \) \\
\hline
\end{tabular}
\caption{Typical mean-field interaction potentials in Rubidium condensates.}
\label{Table_U}
\end{table}

In order to satisfy this combination of requirements, we use typical
parameter values given in Table \ref{STIRAP_param}, with \( U_{11}=4.96\times 10^{-17} \)
m\( ^{3} \)/s. Simulating Eqs. (\ref{J-3}) with these parameter
values and with all other couplings \( U_{ij} \) set to zero, gives
a maximum of \( \eta \simeq 0.95 \) conversion efficiency, for the
optimum delay coefficient of \( \alpha \simeq 1.5 \).

As the next step, one can include the mean-field energies due to atom-molecule
(\( U_{12} \), \( U_{13} \)) and molecule-molecule (\( U_{22} \),
\( U_{23} \), \( U_{33} \)) scattering processes. Considering the
fact that the excited molecular state never gets highly populated
in STIRAP, only processes described by \( U_{12} \) and \( U_{22} \)
are to be taken into account here. Provided that the scattering lengths
for these processes are of the same order of magnitude as the atom-atom
scattering length, these terms do not lead to a dramatic change in
the conversion efficiency.

To account for the most recent experimental data on ultracold atom-molecule
scattering in a \( ^{87} \)Rb condensate \cite{WFHRH2000}, we have
included the \( U_{12} \) term with \( a_{12}=-9.346 \) nm. In addition,
we include the \( U_{22} \) term with an assumption that \( a_{2}=a_{1} \)
and set \( U_{3i}=0 \) since these are currently not known. The resulting
values of \( U_{ij} \) are summarized in Table \ref{Table_U}. 

\begin{figure}
\par\centering \resizebox*{6cm}{!}{\includegraphics{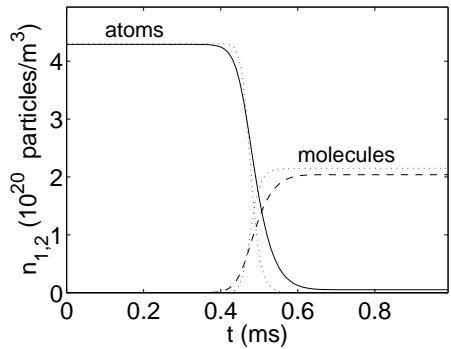} } \par{}
\caption{Efficient conversion of an atomic condensate into a molecular condensate
during STIRAP as obtained by simulating Eqs. (\ref{J-3}), with an
initial atom number density \protect\( n_{1}(0)=4.3\times 10^{20}\protect \)
m\protect\( ^{-3}\protect \). Other parameter values are as in Fig.
2, and Tables \ref{STIRAP_param} and \ref{Table_U}. The solid line
indicates atomic density, the dashed line the molecular density, and
the dotted line the analytic result in the adiabatic limit. }
\label{Conversion93}
\end{figure}

The results of simulations are given in Fig. \ref{Conversion93},
where we see about \( 93\% \) (\( \eta \simeq 0.93 \)) conversion
of the atomic condensate into the condensate of molecules, for \( \Omega _{1}^{(eff,0)}=\Omega _{2}^{(eff,0)}=2.1\times 10^{7} \)
s\( ^{-1} \), \( T=10^{-4} \) s, and \( \alpha \simeq 1.5 \). This
figure also includes the analytic theory calculated in the adiabatic
limit for comparison, and shows that for these parameters, the results
of the numerical simulation are close to those from the adiabatic
theory.

\begin{figure}
\par\centering \resizebox*{8cm}{!}{\includegraphics{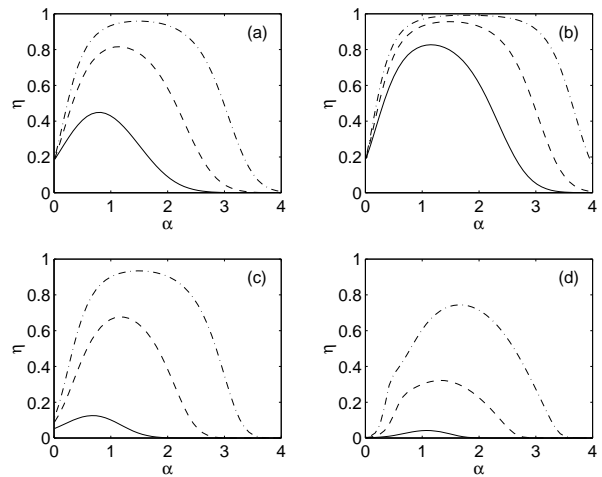} } \par{}
\caption{The conversion efficiency \protect\( \protect \eta \protect \) as
a function of relative delay \protect\( \protect \alpha \protect \),
for: (a) \protect\( T=10^{-4}\, s\protect \) and \protect\( U_{ij}=0\protect \);
(b) \protect\( T=10^{-3}\, s\protect \) and \protect\( U_{ij}=0\protect \);
(c) \protect\( T=10^{-4}\, s\protect \) and \protect\( U_{ij}\protect \)
as in Table \ref{Table_U}; (d) \protect\( T=10^{-3}\, s\protect \)
and \protect\( U_{ij}\protect \) as in Table \ref{Table_U}. The
full, dashed, and dashed-dotted lines correspond to effective Rabi
frequencies \protect\( \Omega _{1}^{(eff,0)}=\Omega _{2}^{(eff,0)}\protect \)
equal to \protect\( 2.1\times 10^{6}\protect \) s\protect\( ^{-1}\protect \),
\protect\( 6.3\times 10^{6}\protect \) s\protect\( ^{-1}\protect \),
and \protect\( 2.1\times 10^{7}\protect \) s\protect\( ^{-1}\protect \).
Other parameter values are as in Table \ref{STIRAP_param}. }
\label{Efficiency-vs-delay}
\end{figure}

Thus, we conclude that even including the mean field energies STIRAP
can be carried out, provided one uses faster time scales than in the
absence of the \( s \)-wave scattering. As a consequence the effective
Rabi frequencies have to be kept at a rather high value. Characteristic
results for comparison are summarized in Fig. \ref{Efficiency-vs-delay},
where we plot conversion efficiency \( \eta  \) versus the relative
delay coefficient \( \alpha  \), for cases where \( s \)-wave scattering
are present or absent, and for different values of the effective Rabi
frequencies and pulse durations \( T \).

Figure \ref{Efficiency-vs-delay} (a) shows a reasonably efficient
conversion, in the absence of mean-field interactions, but including
losses. As expected, spontaneous emission losses are reduced, and
efficiency is improved further by the use of longer pulses, further
into the adiabatic limit, as in Fig. \ref{Efficiency-vs-delay} (b).
However, the more realistic example given in Fig. \ref{Efficiency-vs-delay}
(c) which includes mean-field interactions shows rather poor conversion,
especially when we use smaller effective Rabi frequencies, \( \Omega _{1}^{(eff,0)}=\Omega _{2}^{(eff,0)}=2.1\times 10^{6} \)
s\( ^{-1} \), as shown by the full line where maximum \( \eta \simeq 0.12 \)
at optimum \( \alpha \simeq 0.7 \). This is caused by the effective
two-photon detunings induced by the mean-field interactions. The situation
is made even worse rather than better in Fig. \ref{Efficiency-vs-delay}
(d) when longer pulses are chosen, giving more time for two-photon
detunings to occur.

To be more specific about values of the effective Rabi frequencies
we recall that the definition of \( \Omega _{i}^{(eff)} \) involves
the Franck-Condon overlap integrals \( I_{i,3} \) and {}``bare''
electronic Rabi frequencies \( \bar{\Omega }_{i}^{(el)}=|{\bar{\mathbf{d}}}_{M}\cdot {\mathbf{E}}_{i}|/\hbar  \).
Given the values of \( {\bar{\mathbf{d}}}_{M} \) and \( I_{i,3} \)
which are specific for particular dimer species involved, the size
of \( \Omega _{i}^{(eff,0)} \) can be translated to the intensities
of the Raman lasers. Considering \( ^{87} \)Rb\( _{2} \) as an example,
and using a characteristic values of the corresponding Franck-Condon
integrals, \( |I_{1,3}|\simeq 10^{-14}{\, } \)m\( ^{3/2} \) and
\( |I_{2,3}|\simeq 0.1 \) \cite{HWDK2000}, the magnitudes of \( \Omega _{1}^{(eff,0)}=\Omega _{2}^{(eff,0)}=2.1\times 10^{7} \)
s\( ^{-1} \) translate to peak values of the {}``bare'' Rabi frequencies
equal to \( \Omega _{1}^{(el,0)}=10^{11} \) s\( ^{-1} \) (for \( n_{1}(0)=4.3\times 10^{20} \)
m\( ^{-3} \)) and \( \Omega _{2}^{(el,0)}=2.1\times 10^{8} \) s\( ^{-1} \).
The peak Rabi frequency of \( \Omega _{1}^{(el,0)}=10^{11} \) s\( ^{-1} \)
for the free-bound transition would be realized with a \( 1 \) W
laser power and a waist size of about 10 \( \mu  \)m, which is not
impossible -- but much higher than we would estimate without the combined
effects of spontaneous emission and collisional processes. Another
obvious problem here is that the waist size of \( 10 \) \( \mu  \)m
is comparable to characteristic spatial extend of current BECs in
a trap.

In summary, our analysis shows that the relatively small overlap integrals
for the free-bound transitions, together with the mean-field interaction
detunings, can require rather high intensity of the \( \omega _{1} \)-laser
for obtaining high conversion efficiencies.

\subsection{Off-resonance operation}

In order to allow one to operate under less demanding laser powers
or \emph{smaller} Rabi frequencies (e.g. \( \Omega _{1}^{(eff,0)}=\Omega _{2}^{(eff,0)}=2.1\times 10^{6} \)
s\( ^{-1} \)) -- while still maintaining efficient conversion --
we now consider the role of the detunings \( \Delta _{1} \) and \( \Delta _{2} \)
in the off-resonance regime of operation. In effect, this approach
relies on compensating for the phase shifts due to the mean field
energies, and tuning the free-bound and bound-bound transitions to
a {}``true'' resonance. The physics behind this is that in BEC environments
it is not appropriate to consider transitions with respect to single-particle
bare energies \( E_{i} \). Rather, the relevant energies and therefore
the effective resonances have to take into account the mean-field
energy contributions due to self- and cross-interactions between the
condensates.

\begin{figure}
\par\centering \resizebox*{6cm}{!}{\includegraphics{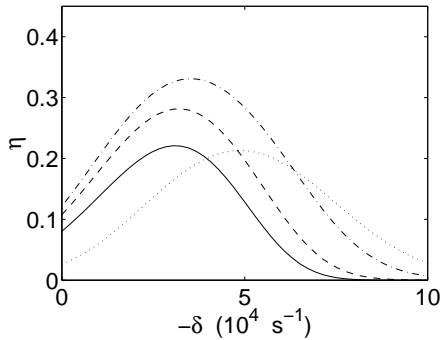} } \par{}
\caption{The conversion efficiency \protect\( \protect \eta \protect \) as
a function of \protect\( -\protect \delta \protect \), for \protect\( \Omega ^{(eff,0)}_{1}=\Omega ^{(eff,0)}_{2}=2.1\times 10^{6}\protect \)
s\protect\( ^{-1}\protect \), \protect\( T=10^{-4}\protect \) s
and different delay coefficients: \protect\( \protect \alpha =0.2\protect \)
(full line), \protect\( \protect \alpha =0.4\protect \) (dashed line),
\protect\( \protect \alpha =0.8\protect \) (dashed-dotted line),
\protect\( \protect \alpha =1.5\protect \) (dots). Other parameter
values are as in Tables \ref{STIRAP_param} and \ref{Table_U}. }
\label{Efficiency-vs-delta}
\end{figure}

More specifically, it is the \emph{two-photon} detuning \( \delta  \)
that has to be adjusted to the \emph{relative} phase between the atomic
and molecular condensates. Alternatively speaking, by tuning the two-photon
detuning to compensate for the net mean-field energy, one reduces
the effect of dephasing since the effective dephasing time becomes
longer compared to the pulse durations. The problem, however, is more
complicated because the mean field energy is changed dynamically as
the populations of the atomic and the molecular condensate themselves
are being changed during STIRAP. As a crude estimate of an appropriate
value of \( \delta  \) one can simply choose it to compensate the
\emph{initial} mean field energy in the atomic condensate. This approach
-- employed for smaller Rabi frequencies than before -- can substantially
improve the conversion efficiency, compared to the case of zero two-photon
detuning.

To show this we have carried out simulations with ten times smaller
Rabi frequencies than before (i.e. with \( \Omega _{i}^{(eff,0)}=2.1\times 10^{6} \)
s \( ^{-1} \)), corresponding to a decrease of the Raman laser intensities
by a factor of \( 100 \). The results are summarized in Fig. \ref{Efficiency-vs-delta}
, where we plot the conversion efficiency \( \eta  \) versus \( \delta  \),
for \( T=10^{-4} \) s and different delay coefficients \( \alpha  \).

As we see, by varying the two-photon detuning and tuning it to the
optimum value one can improve the conversion efficiency by about a
factor of two or more, for a range of values of the delay coefficient.
Furthermore, as the effective dephasing time is increased when the
contribution of the mean field energies is compensated by \( \delta  \),
one can further improve the results by employing longer pulse durations.
More generally, the problem of finding a set of values of \( T \),
\( \alpha  \), and \( \delta  \) that maximize the conversion efficiency,
for given values of the effective Rabi frequencies, is now transformed
to an optimization problem.

\subsection{Asymmetric effective Rabi frequencies}

We now wish to explore an alternative strategy for improving the conversion
efficiency under experimentally less demanding conditions of smaller
Rabi frequencies. We consider the effects of non-equal effective Rabi
frequencies.

Using the earlier given characteristic values of the Franck-Condon
overlap integrals, \( |I_{1,3}|\simeq 10^{-14}{\, } \)m\( ^{3/2} \)
and \( |I_{2,3}|\simeq 0.1 \) , we can estimate that the moderate
magnitudes of \( \Omega _{1}^{(eff,0)}=\Omega _{2}^{(eff,0)}=2.1\times 10^{6} \)
s\( ^{-1} \) translate to the following peak values of the bare Rabi
frequencies: \( \Omega _{1}^{(el,0)}=10^{10} \) s\( ^{-1} \) (for
\( n_{1}(0)\sim 4.3\times 10^{20} \) m\( ^{-3} \)) and \( \Omega _{2}^{(el,0)}=2.1\times 10^{7} \)
s\( ^{-1} \). As we see, while this corresponds to equal \emph{effective}
Rabi frequencies, however, the absolute values of the corresponding
\emph{bare} Rabi frequencies are \emph{not} equal. The limitation
on laser intensities refers primarily to the free-bound transition,
whose bare Rabi frequency \( \Omega _{1}^{(el,0)} \) is higher.

\begin{figure}
\par\centering \resizebox*{6cm}{!}{\includegraphics{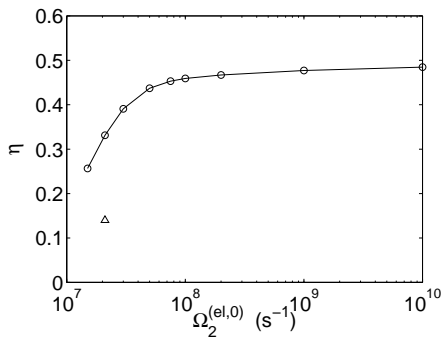} } \par{}
\caption{The maximum conversion efficiency \protect\( \protect \eta \protect \)
versus \protect\( \Omega ^{(el,0)}_{2}\protect \) (the evaluated
points are represented by circles), for \protect\( \Omega ^{(el,0)}_{1}=10^{10}\protect \)
s\protect\( ^{-1}\protect \) and the corresponding optimum values
of \protect\( T\protect \), \protect\( \protect \alpha \protect \),
and \protect\( \protect \delta \protect \) as given in Table \ref{Table_main}.
The triangle gives the result of an optimization with \protect\( \protect \delta =0\protect \). }
\label{Optimum-efficiency}
\end{figure}

As far as the second Rabi frequency \( \Omega _{2}^{(el,0)} \) is
concerned, one can in principle increase its magnitude up to the same
value as \( \Omega _{1}^{(el,0)} \), i.e. \( \Omega _{2}^{(el,0)}=10^{10} \)
s\( ^{-1} \), thus maintaining experimentally similar and reasonably
high intensities for both lasers. Under these conditions, and for
the same values of the Franck-Condon overlap integrals and \( n_{1}(0) \),
we would have \begin{eqnarray}
\Omega _{1}^{(el,0)} & = & 10^{10}\text {s}^{-1}\text {{}},\; \; \Omega _{1}^{(eff,0)}=2.1\times 10^{6}\text {s}^{-1},\nonumber \\
\Omega _{2}^{(el,0)} & = & 10^{10}\text {s}^{-1},\text {\thinspace \, \, \thinspace \, \, }\Omega _{2}^{(eff,0)}=10^{9}\text {s}^{-1}.\label{Rabi-unequal} 
\end{eqnarray}

We can now ask the question of what happens in STIRAP with \emph{different}
effective Rabi frequencies, and whether one can achieve higher conversion
efficiencies in the regime where \( \Omega _{2}^{(eff,0)}\gg \Omega _{1}^{(eff,0)} \).
This approach again leads to an increased conversion efficiency compared
to the case of \emph{equal} effective Rabi frequencies. To generalize
the analysis, we now treat different cases as an optimization problem
(that maximize \( \eta  \)), carried out with respect to \( T \),
\( \alpha  \), and \( \delta  \), for a set of different values
of \( \Omega _{2}^{(el,0)} \) within a range of \( \Omega _{2}^{(el,0)}=1.5\times 10^{7}-10^{10} \)
s\( ^{-1} \), and for a given value of \( \Omega _{1}^{(el,0)}=10^{10} \)
s\( ^{-1} \). In terms of the effective Rabi frequencies, this corresponds
to \( \Omega _{2}^{(eff)} \) ranging from \( 1.5\times 10^{6} \)
to \( 10^{9} \) s\( ^{-1} \), for a given \( \Omega _{1}^{(eff,0)}=2.1\times 10^{6} \)
s\( ^{-1} \).

The results are summarized in Fig. \ref{Optimum-efficiency} and in
Table \ref{Table_main} where we give the corresponding optimum values
of \( T \), \( \alpha  \), and \( \delta  \), and the resulting
maximum conversion efficiency \( \eta  \). 

For comparison, in the symmetric case of \( \Omega _{1}^{(eff,0)}=\Omega _{2}^{(eff,0)}=2.1\times 10^{6} \)
s\( ^{-1} \) and \( \delta =0 \), such that the optimization is
carried out only with respect to \( T \) and \( \alpha  \), the
maximum conversion efficiency would be \( \eta \simeq 0.14 \) (at
optimum \( T=0.46\times 10^{-4} \) s and \( \alpha =0.54 \)). This
case is represented by the triangle, in Fig. \ref{Optimum-efficiency}.

\begin{table}
\begin{tabular}{|c|c|c|c|c|}
\hline 
\( \Omega _{2}^{(el,0)} \) (s\( ^{-1} \)) &
 \( T \) (\( \times 10^{-4} \) s) &
 \( \alpha  \)&
 \( \delta  \) (\( \times 10^{4} \) s\( ^{-1} \)) &
 \( \eta  \)\\
\hline
\( 1.5\times 10^{7} \)&
 \( 0.987 \)&
 \( 0.753 \)&
 \( 3.58 \)&
 \( 0.257 \)\\
\hline
\( 2.1\times 10^{7} \)&
 \( 0.966 \)&
 \( 0.795 \)&
 \( 3.57 \)&
 \( 0.331 \)\\
\hline
\( 3\times 10^{7} \)&
 \( 1.05 \)&
 \( 0.882 \)&
 \( 3.44 \)&
 \( 0.391 \)\\
\hline
\( 5\times 10^{7} \)&
 \( 1.29 \)&
 \( 1.05 \)&
 \( 3.11 \)&
 \( 0.437 \)\\
\hline
\( 7.5\times 10^{7} \)&
 \( 1.49 \)&
 \( 1.19 \)&
 \( 2.92 \)&
 \( 0.453 \)\\
\hline
\( 10^{8} \)&
 \( 1.61 \)&
 \( 1.30 \)&
 \( 2.83 \)&
 \( 0.459 \)\\
\hline
\( 2\times 10^{8} \)&
 \( 1.86 \)&
 \( 1.53 \)&
 \( 2.71 \)&
 \( 0.467 \)\\
\hline
\( 10^{9} \)&
 \( 2.32 \)&
 \( 1.98 \)&
 \( 2.59 \)&
 \( 0.474 \)\\
\hline
\( 10^{10} \)&
 \( 2.58 \)&
 \( 2.51 \)&
 \( 2.93 \)&
 \( 0.486 \) \\
\hline
\end{tabular}
\caption{Optimum STIRAP parameters for: \protect\( \Omega _{1}^{(el,0)}=10^{10}\text {s}^{-1}\protect \),
\protect\( |I_{1,3}|=10^{-14}\protect \) m\protect\( ^{3/2}/s\protect \),
and \protect\( n_{1}(0)=4.3\times 10^{20}\protect \) m\protect\( ^{-3}\protect \),
so that \protect\( \Omega _{1}^{(eff,0)}=2.1\times 10^{6}\text {s}^{-1}\protect \)
in all cases; different values of \protect\( \, \Omega _{2}^{(eff,0)}=I_{1,3}\Omega _{2}^{(el,0)}\protect \)
are taken for \protect\( |I_{1,3}|=0.1\protect \) and \protect\( \Omega _{2}^{(el,0)}\protect \)
ranging from \protect\( 1.5\times 10^{7}\protect \) to \protect\( 10^{10}\protect \)
s\protect\( ^{-1}\protect \).}
\label{Table_main}
\end{table}

Thus, we have shown that by introducing the possibility of varying
the two-photon detuning \( \delta  \) and the Rabi frequency \( \Omega _{2}^{(el,0)} \),
the conversion efficiency can be increased almost by a factor of \( 4 \).
This can be crucial for experimental observation of the phenomenon
of coherent conversion of an atomic BEC into a molecular BEC, via
STIRAP.

\section{Realistic condensate models}

\subsection{Uniform multi-level model}

In our model for STIRAP we only treated the coupling of laser \( \omega _{1} \)
to the free-bound transition \( \left| 1\right\rangle \leftrightarrow \left| 3\right\rangle  \)
with Rabi frequency \( \Omega _{1}=\bar{\Omega }_{1}^{(el)}I_{1,3} \)
together with the coupling of laser \( \omega _{2} \) to the bound-bound
transition \( \left| 2\right\rangle \leftrightarrow \left| 3\right\rangle  \)
with Rabi frequency \( \Omega _{2}=\bar{\Omega }_{2}^{(el)}I_{2,3} \).
This approximation can only be valid if the laser \( \omega _{1} \)
is far detuned from the \( \left| 2\right\rangle \leftrightarrow \left| 3\right\rangle  \)
transition, and similarly -- if the laser \( \omega _{2} \) is far
detuned from the \( \left| 1\right\rangle \leftrightarrow \left| 3\right\rangle  \)
transition. In addition, the two lasers have to be far detuned from
transitions to any other vibrational levels \( \left| v^{\prime }\right\rangle  \)
(adjacent to \( \left| 3\right\rangle  \)) in the excited potential.
We define the relevant detunings, for the simplest uniform case, as
follows: \begin{eqnarray}
\Delta _{13,\omega _{2}} & = & (E_{3}-2E_{1})/\hbar -\omega _{2},\nonumber \\
\Delta _{23,\omega _{1}} & = & (E_{3}-E_{2})/\hbar -\omega _{1},\nonumber \\
\Delta _{1v^{\prime },\omega _{1}} & = & (E_{v^{\prime }}-2E_{1})/\hbar -\omega _{1},\nonumber \\
\Delta _{1v^{\prime },\omega _{2}} & = & (E_{v^{\prime }}-2E_{1})/\hbar -\omega _{2},\nonumber \\
\Delta _{2v^{\prime },\omega _{1}} & = & (E_{v^{\prime }}-E_{2})/\hbar -\omega _{1},\nonumber \\
\Delta _{2v^{\prime },\omega _{2}} & = & (E_{v^{\prime }}-E_{2})/\hbar -\omega _{2}.
\end{eqnarray}

In general, these cross-couplings -- if included into the model --
lead to incoherent radiative losses of atoms and molecules due to
spontaneous emission, which modifies the effective detunings to:
\begin{equation}
 \Delta ^{\gamma }_{iv^{\prime },\omega _{j}}=\Delta _{iv^{\prime },\omega _{j}}+i\gamma /2  \, .
 \end{equation}
In order that these losses be negligible we require the respective
detunings to be large enough. This requirement, however, may not be
easily satisfied, as the magnitudes of the detunings are in principle
limited from above by the characteristic distance between the adjacent
vibrational levels of the excited molecular potential. In other words,
increasing the detuning with respect to one transition will eventually
bring the laser frequency to a resonance with respect to the nearby
level. More importantly, these cross-couplings provide scattering
pathways that are not cancelled out in a dark-state interference effect,
so that their overall disruptive effect -- over the adiabatically
long pulse durations -- may turn out to be rather large.

In order to estimate these effects, we therefore explicitly include
all other relevant coupling processes into our model. In addition
to losses, the incoherent couplings induce light shifts that effectively
lead to a dephasing between the atomic and molecular condensates.
Treating these, leads to the following additional terms in the STIRAP
equations, in same rotating frames as in Eqs. (\ref{TF}): \begin{eqnarray}
\frac{\partial \psi _{1}}{\partial t} & = & (...\, )+i\beta ^{\gamma }_{1}\psi _{1}+i\bar{U}^{\gamma }_{11}\left| \psi _{1}\right| ^{2}\psi _{1}\nonumber \\
 &  & -i\chi \psi _{1}^{\ast }\psi _{2}+i\frac{({\bar{\Omega }}_{2}^{(el)}I_{1,3})^{\ast }}{\sqrt{2}}e^{-i\omega _{12}t}\psi _{1}^{\ast }\psi _{3},\label{losses1} 
\end{eqnarray}
\begin{eqnarray}
\frac{\partial \psi _{2}}{\partial t} & = & (...\, )+i\beta ^{\gamma }_{2}\psi _{2}\nonumber \\
 &  & -i\frac{\chi ^{'}}{2}\psi _{1}^{2}+i\frac{({\bar{\Omega }}_{1}^{(el)}I_{2,3})^{\ast }}{2}e^{i\omega _{12}t}\psi _{3},\label{losses2} 
\end{eqnarray}
\begin{eqnarray}
\frac{\partial \psi _{3}}{\partial t} & = & (...\, )+i\frac{{\bar{\Omega }}_{2}^{(el)}I_{1,3}}{2\sqrt{2}}e^{i\omega _{12}t}\psi _{1}^{2}\nonumber \\
 &  & +i\frac{{\bar{\Omega }}_{1}^{(el)}I_{2,3}}{2}e^{-i\omega _{12}t}\psi _{2},\label{losses3} 
\end{eqnarray}
 where \( (...\, ) \) stand for the terms already present in the
right hand sides of Eqs. (\ref{TF}), and
 \( \omega _{12}=\omega _{1}-\omega _{2}, \) . Here,
 the effective  complex light shift coefficients \( \beta ^{\gamma }_{1} \)
and \( \beta ^{\gamma }_{2} \) , the nonlinear phase shift \( \bar{U}_{11} \)
(which effectively leads to a modified atom-atom scattering length)
and the effective parametric couplings \( \chi  \), \( \chi ' \)
(including only non-oscillating terms) are given by:\begin{equation}
\beta ^{\gamma }_{1}=\frac{|\Omega _{1}^{(A)}|^{2}}{4D^{\gamma }_{1}}+\frac{|\Omega _{2}^{(A)}|^{2}}{4D^{\gamma }_{2}},
\end{equation}
\begin{equation}
\beta ^{\gamma }_{2}= \sum\nolimits _{v^{\prime }}^{\prime }\left [ \frac{|{\bar{\Omega }}_{1}^{(el)}I_{2,v^{\prime }}|^{2}}{4\Delta ^{\gamma }_{2v^{\prime },\omega _{1}}}+\frac{|{\bar{\Omega }}_{2}^{(el)}I_{2,v^{\prime }}|^{2}}{4\Delta ^{\gamma }_{2v^{\prime },\omega _{2}}}\right] ,
\end{equation}
\begin{equation}
\bar{U}^{\gamma }_{11}=\sum\nolimits _{v^{\prime }}^{\prime }\left[ \frac{|{\bar{\Omega }}_{1}^{(el)}I_{1,v^{\prime }}|^{2}}{4\Delta ^{\gamma }_{1v^{\prime },\omega _{1}}}+\frac{|{\bar{\Omega }}_{2}^{(el)}I_{1,v^{\prime }}|^{2}}{4\Delta ^{\gamma }_{1v^{\prime },\omega _{2}}}\right] ,
\end{equation}
\begin{equation}
\label{chi}
\chi =-\frac{{\bar{\Omega }}_{1}^{(el)\ast }{\bar{\Omega }}_{2}^{(el)}}{2\sqrt{2}}\sum\nolimits _{v^{\prime }}^{\prime }\frac{I^{\ast }_{1,v^{\prime }}I_{2,v^{\prime }}}{\Delta ^{\gamma }_{1v^{\prime },\omega _{1}}}.
\end{equation}
\begin{equation}
\label{chi'}
\chi '=-\frac{{\bar{\Omega }}_{1}^{(el)}{\bar{\Omega }}_{2}^{(el)\ast }}{2\sqrt{2}}
\sum\nolimits _{v^{\prime }}^{\prime }\frac{I_{1,v^{\prime }}I^{\ast }_{2,v^{\prime }}}{\Delta ^{\gamma }_{1v^{\prime },\omega _{1}}}.
\end{equation}
where \( D_{1}=\omega _{0}-\omega _{1} \) and \( D_{2}=\omega _{0}-\omega _{2} \)
represent the detunings of lasers \( \omega _{1} \) and \( \omega _{2} \)
from the resonance frequency \( \omega _{0} \) of the atomic transition
between the dissociation limits of the ground and excited potentials.
In addition, 
\begin{equation}
 D^{\gamma }_{j}=D_{j}+i\gamma _{A}/2 \, ,
 \end{equation}
where
\( \gamma _{A}=\gamma /2 \) is the atomic spontaneous decay rate,
and \( \Omega _{1}^{(A)} \) and \( \Omega _{2}^{(A)} \) are the
atomic Rabi frequencies which we take \( \Omega _{i}^{(A)}= \) \( \overline{\Omega }_{i}^{(el)}/\sqrt{2} \).

We now introduce real frequency shift and loss coefficients, so that
\( \beta ^{\gamma }_{1}=\beta _{1}+i\alpha /2 \) , \( \beta ^{\gamma }_{2}=\beta _{1}+i\Gamma _{2}/2 \)
and \( \bar{U}^{\gamma }_{11}=\bar{U}_{11}+i\Gamma _{1}/2 \) . Note
that, in general, \( \chi ^{\ast }\neq \chi ' \) . This means that
the parametric terms are not population-preserving, and can provide
a STIRAP-type of loss reduction even for the non-resonant vibrational
levels, provided the coupling is STIRAP-like.

\begin{figure}
\par\centering \resizebox*{4.5cm}{!}{\includegraphics{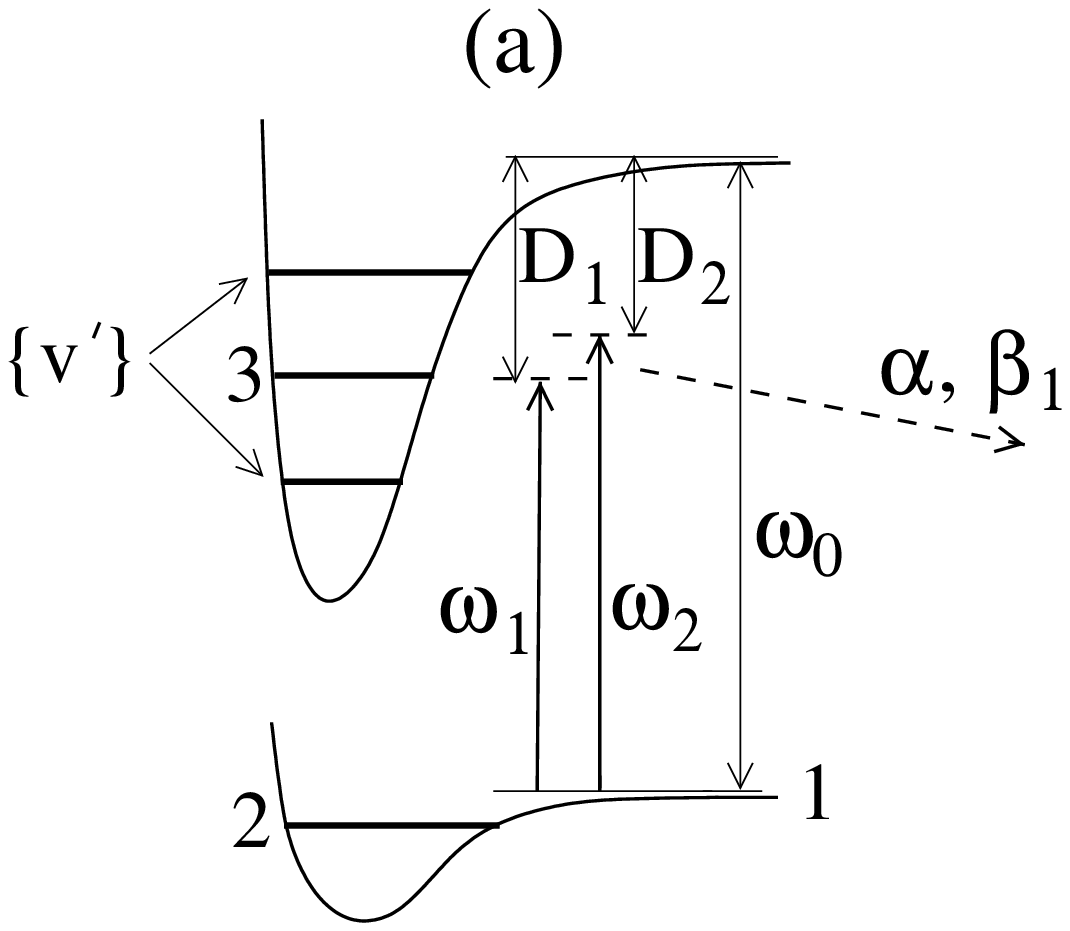} } \par{} \bigskip
\par\centering \resizebox*{4cm}{!}{\includegraphics{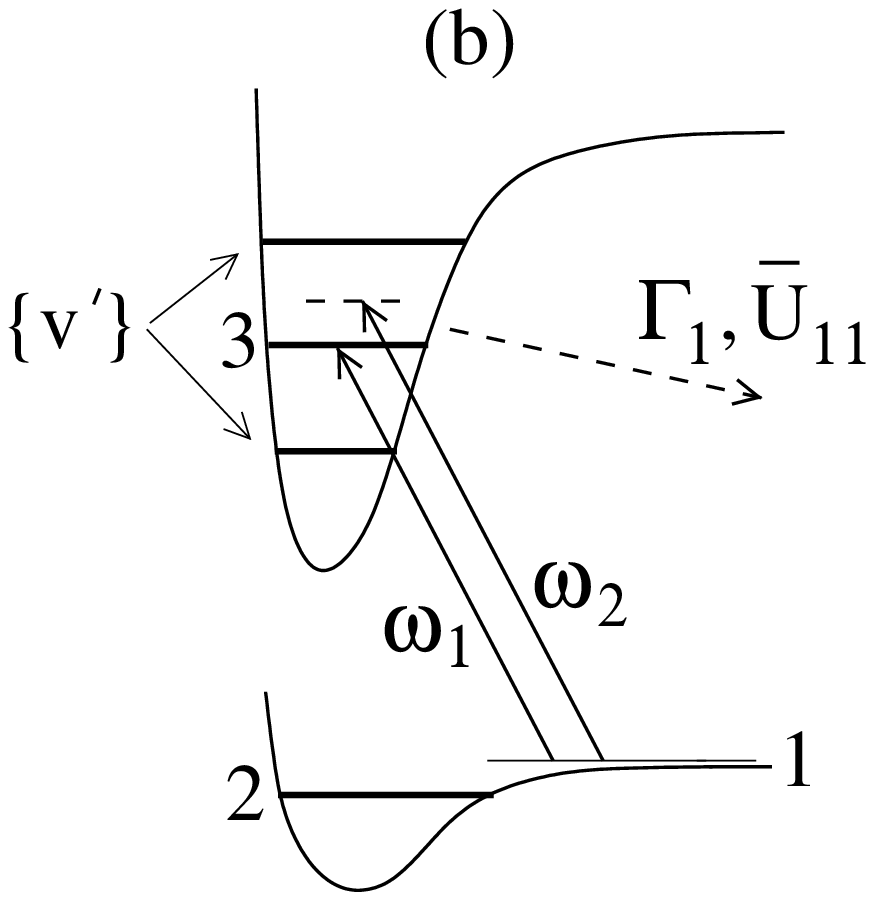} } \par{} \bigskip
\par\centering \resizebox*{4cm}{!}{\includegraphics{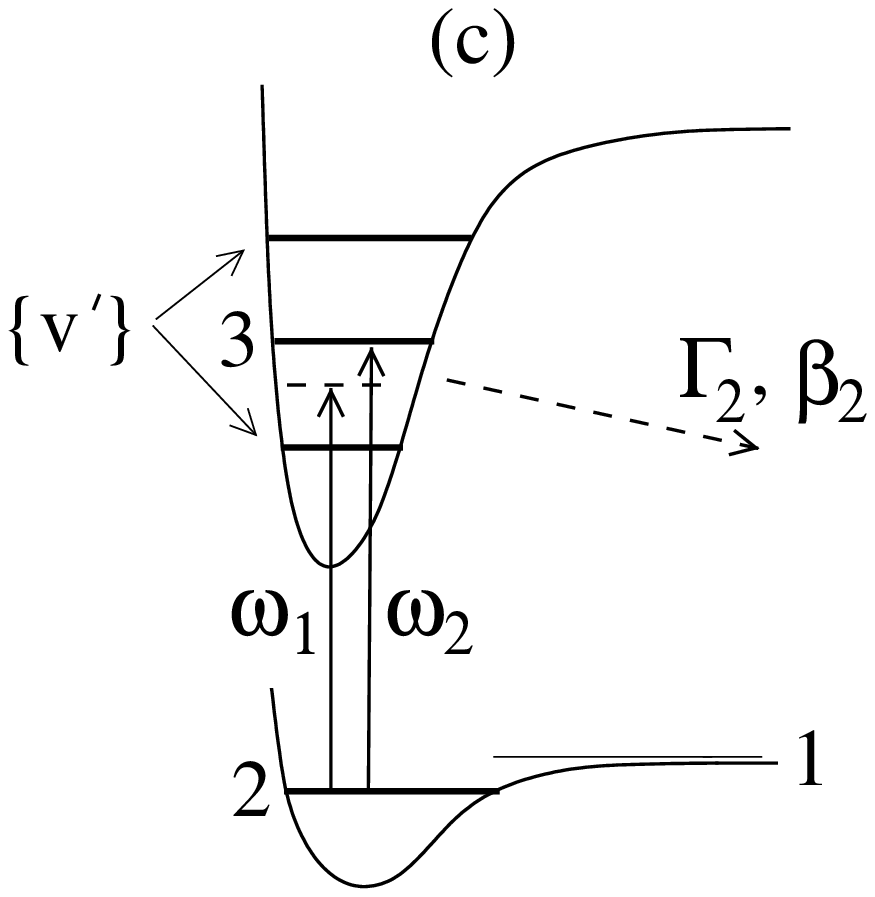} } \par{} \bigskip
\caption{Diagramatic representation of incoherent scattering processes resulting
in induced losses and light shifts.}
\label{Incoherent processes}
\end{figure}

The coefficients \( \alpha  \), \( \Gamma _{i} \), \( \beta _{i} \),
\( \bar{U}_{11} \)\ and \( \chi  \) are obtained by explicitly treating
all other levels in the excited potential, adjacent to \( \left| 3\right\rangle  \),
followed by the procedure of adiabatic elimination. The coefficients
also include the contributions from Raman type of couplings \( \left| 1\right\rangle \leftrightarrow \left| v^{\prime }\right\rangle  \)
by the \( \omega _{1} \)-laser and \( \left| 2\right\rangle \leftrightarrow \left| v^{\prime }\right\rangle  \)
by \( \omega _{2} \)-laser. In principle, these additional Raman
couplings could be treated exactly like the primary STIRAP transition
via \( \left| 3\right\rangle  \), i.e. taking place via the dark-state
interference effect, except that the transitions have much larger
one-photon detuning. This would require an adiabaticity condition
of the form of Eq. (\ref{adiabatic-1}) that includes the one-photon
detuning, implying that a larger value of the product \( \Omega _{i}^{(eff,0)}T \)
is needed. However, \( \Omega _{i}^{(eff,0)}T \)\ \ can not be made
arbitrarily large, as we discussed earlier. Therefore our approach
is to treat these extra Raman couplings as loss and dephasing processes,
rather than to include them into the adiabatic passage scheme. The
contribution of these couplings to the effective atom-molecule conversion
rate, described by \( \chi  \), is negligibly small compared to the
conversion rate due to the parimary Raman transition via \( \left| 3\right\rangle  \).

The relevant transitions that stand behind these coefficients are
illustrated in Fig. \ref{Incoherent processes}. For example, the
coefficient $\alpha $ describes the process of atomic absorption
from either of the two Raman lasers, that incoherently produce excited
atoms followed by spontaneous emission loss. The coefficient $\Gamma _{1}$
is due to ordinary photoassociation when pairs of atoms from the condensate
are transferred (again by either of the two Raman lasers) into an
excited molecular state which can then spontaneously dissociate into
a pair of hot (non-condensed) atoms. The effective rate of this non-linear
loss is $\Gamma _{1}n_{1}$. Finally, the coefficient $\Gamma _{2}$
describes 
the loss of molecules due to spontaneous Raman scattering of laser photons.
This produces molecules predominantly in ro-vibrational levels other than the
one targetted by the stimulated Raman transition. This term may also describe some scattering
which is elastic in the sense that the molecules return to
the targeted state, but with an increased kinetic energy due to photon recoil,
that will still remove the molecules from the condensate.

The summations in the expressions for \( \Gamma _{i} \), \( \beta _{2} \),
\( \bar{U}_{11} \) and \( \chi  \) are carried out over all the
excited levels \( v^{\prime } \) except the resonant level \( \left| 3\right\rangle  \)
which itself participates in STIRAP, rather than being adiabatically
eliminated. The effects of losses and light shifts due to the cross-couplings
to the level \( \left| 3\right\rangle  \) itself are implicitly described
by the last terms in the rhs of Eqs. (\ref{losses1})-(\ref{losses3}).
Subsequently, we will estimate the combined effects of all levels
in which case the contribution of the level \( \left| 3\right\rangle  \)
is estimated by similar terms to the ones included in \( \Gamma _{i} \),
\( \beta _{2} \), \( \bar{U}_{11} \),\ except that the detunings
\( \Delta _{1v\prime ,\omega _{2}} \) and \( \Delta _{2v\prime ,\omega _{1}} \)
are replaced by \( \Delta _{13,\omega _{2}} \) and \( \Delta _{23,\omega _{1}} \),
respectively, and \( I_{i,v^{\prime }} \) are replaced by \( I_{i,3} \).

Separating out the time dependences of the two Rabi frequencies, the
above coefficients can be rewritten as: \begin{equation}
\alpha =\alpha ^{(1)}e^{-2(t-t_{1})^{2}/T^{2}}+\alpha ^{(2)}e^{-2(t-t_{2})^{2}/T^{2}},
\end{equation}
\begin{equation}
\Gamma _{i}=\Gamma _{i}^{(1)}e^{-2(t-t_{1})^{2}/T^{2}}+\Gamma _{i}^{(2)}e^{-2(t-t_{2})^{2}/T^{2}},
\end{equation}
\begin{equation}
\beta _{i}=\beta _{i}^{(1)}e^{-2(t-t_{1})^{2}/T^{2}}+\beta _{i}^{(2)}e^{-2(t-t_{2})^{2}/T^{2}},
\end{equation}
\begin{equation}
\bar{U}_{11}=\bar{U}_{11}^{(1)}e^{-2(t-t_{1})^{2}/T^{2}}+\bar{U}_{11}^{(2)}e^{-2(t-t_{2})^{2}/T^{2}},
\end{equation}
\begin{equation}
\chi =\chi _{0}e^{-(t-t_{1})^{2}/T^{2}}e^{-(t-t_{2})^{2}/T^{2}},
\end{equation}
 where the peak values are, respectively: \begin{equation}
\alpha ^{(i)}=\frac{\gamma _{A}}{4}\left| \frac{\Omega _{i}^{(A,0)}}{D_{i}}\right| ^{2},
\end{equation}
\begin{equation}
\Gamma _{1}^{(i)}=\frac{\gamma }{4}\sum\nolimits _{v^{\prime }}^{\prime }\left| \frac{{\bar{\Omega }}_{i}^{(el,0)}I_{1,v^{\prime }}}{\Delta _{1v^{\prime },\omega _{i}}}\right| ^{2},
\end{equation}
\begin{equation}
\Gamma _{2}^{(i)}=\frac{\gamma }{4}\sum\nolimits _{v^{\prime }}^{\prime }\left| \frac{{\bar{\Omega }}_{i}^{(el,0)}I_{2,v^{\prime }}}{\Delta _{2v^{\prime },\omega _{i}}}\right| ^{2},
\end{equation}
\begin{equation}
\beta _{1}^{(i)}=\frac{|\Omega _{i}^{(A,0)}|^{2}}{4D_{i}},
\end{equation}
\begin{equation}
\beta _{2}^{(i)}=\sum\nolimits _{v^{\prime }}^{\prime }\frac{|{\bar{\Omega }}_{i}^{(el,0)}I_{2,v^{\prime }}|^{2}}{4\Delta _{2v^{\prime },\omega _{i}}},
\end{equation}
\begin{equation}
\bar{U}_{11}^{(i)}=\sum\nolimits _{v^{\prime }}^{\prime }\frac{|{\bar{\Omega }}_{i}^{(el,0)}I_{1,v^{\prime }}|^{2}}{4\Delta _{1v^{\prime },\omega _{i}}}.
\end{equation}

\begin{equation}
\chi _{0}'=\chi _{0}^*=-\frac{{\bar{\Omega }}_{1}^{(el,0)}{\bar{\Omega }}_{2}^{(el,0)\ast }}{2\sqrt{2}}\sum\nolimits _{v^{\prime }}^{\prime }\frac{I_{1,v^{\prime }}I_{2,v^{\prime }}^{\ast }}{\Delta ^{\gamma }_{1v^{\prime },\omega _{1}}}.
\end{equation}

The reason for this separation is that the two terms in each coefficient
act during different time intervals, corresponding to the first and
the second pulse in STIRAP. Accordingly, one has to distinguish their
disruptive effect during the duration of the corresponding pulses.
For example the molecule loss term \( \Gamma _{2}^{(2)} \) acts during
the first Raman pulse (of frequency \( \omega _{2} \)) when the molecular
field is not populated yet. As a result, the coefficient \( \Gamma _{2}^{(2)} \)
is not so disruptive. On the other hand, the molecule loss term \( \Gamma _{2}^{(1)} \)
is much more important since it acts during the second Raman pulse
(with frequency \( \omega _{1} \)) when the population of the molecular
condensate becomes high. If the value of \( \Gamma _{2}^{(1)} \)
is too large, one can easily lose all this population during the \( \omega _{1} \)-pulse.

In order that the radiative losses and dephasing due to light shifts
be negligible over the duration of STIRAP, the time scales associated
with the coefficients \( \alpha  \), \( \Gamma _{1}n_{1} \), \( \Gamma _{2} \),
and the induced \emph{relative} phases must be much larger than the
duration of pulses in STIRAP, i.e. \begin{equation}
\label{condition-alpha}
\left[ \alpha ^{(i)}\right] ^{-1}\gg T,
\end{equation}
\begin{equation}
\left[ \Gamma _{1}^{(1)}n_{1}\right] ^{-1}\gg T,
\end{equation}
\begin{equation}
\left[ \left( \Gamma _{1}^{(2)}+\frac{\gamma }{4}\left| \frac{{\bar{\Omega }}_{2}^{(el,0)}I_{1,3}}{\Delta _{13,\omega _{2}}}\right| ^{2}\right) n_{1}\right] ^{-1}\gg T,
\end{equation}
\begin{equation}
\label{condition-Gamma-2(1)}
\left[ \Gamma _{2}^{(1)}+\frac{\gamma }{4}\left| \frac{{\bar{\Omega }}_{1}^{(el,0)}I_{2,3}}{\Delta _{23,\omega _{1}}}\right| ^{2}\right] ^{-1}\gg T,
\end{equation}
\begin{equation}
\left[ \Gamma _{2}^{(2)}\right] ^{-1}\gg T,
\end{equation}
\begin{equation}
\left| \left( \beta _{2}^{(1)}+\frac{|{\bar{\Omega }}_{1}^{(el,0)}I_{2,3}|^{2}}{4\Delta _{23,\omega _{1}}}\right) -2\beta _{1}^{(1)}\right| ^{-1}\gg T,
\end{equation}
\begin{equation}
\label{condition-beta-mismatch-2}
\left| \beta _{2}^{(2)}-2\beta _{1}^{(2)}\right| ^{-1}\gg T.
\end{equation}
 This will guarantee that the pulses are switched off before the losses
and dephasings can have their disruptive effect. The influence of
the nonlinear phase shift due to \( \bar{U}_{11}^{(i)} \) can be
ignored simply on the grounds of \( \bar{U}_{11}^{(i)}\ll U_{11} \)
which is the case we encounter in our analysis.

In the above conditions involving the coefficients \( \Gamma _{1}^{(2)} \),
\( \Gamma _{2}^{(1)} \), and \( \beta _{2}^{(1)} \), we have included
additional terms which are the contributions from the incoherent cross-couplings
\( 1\leftrightarrow 3 \) by the laser \( \omega _{2} \) and \( 2\leftrightarrow 3 \)
by the laser \( \omega _{1} \). As we mentioned earlier, these processes
are treated explicitly by the last terms in the rhs of Eqs. (\ref{losses1})-(\ref{losses3}).
However, their overall effect can be described by expressions similar
to the corresponding terms in the coefficients \( \Gamma _{1}^{(2)} \),
\( \Gamma _{2}^{(1)} \), and \( \beta _{2}^{(1)} \). Therefore these
additional contributions must be included in the above conditions,
as they play an important role for correct estimates of the overall
degree of disruption due to incoherent couplings.

Our goal now is in performing a realistic analysis of the above coefficients
for the \( ^{87} \)Rb\( _{2} \) molecule under consideration, and
in finding appropriate target levels for the Raman transitions in
STIRAP so that the disruptive effects are minimized. This is done
using the results of calculation \cite{Roahn Wynar  Verhaar} of the
dipole matrix elements, energy eigenvalues, and the Franck-Condon
overlap integrals for model potentials that closely approximate the
\( ^{87} \)Rb\( _{2} \) ground \( ^{3}\sum _{u}^{+} \) potential
and the O\( _{g}^{-} \) symmetry excited potential. The calculation
treats 205 ro-vibrational levels in the excited potential (which we
label by \( v^{\prime }=0,1,2,...,204 \)) and 39 levels (\( v=0,1,2,...,38 \))
in the ground potential.

\begin{figure}
\par\centering \resizebox*{6cm}{!}{\includegraphics{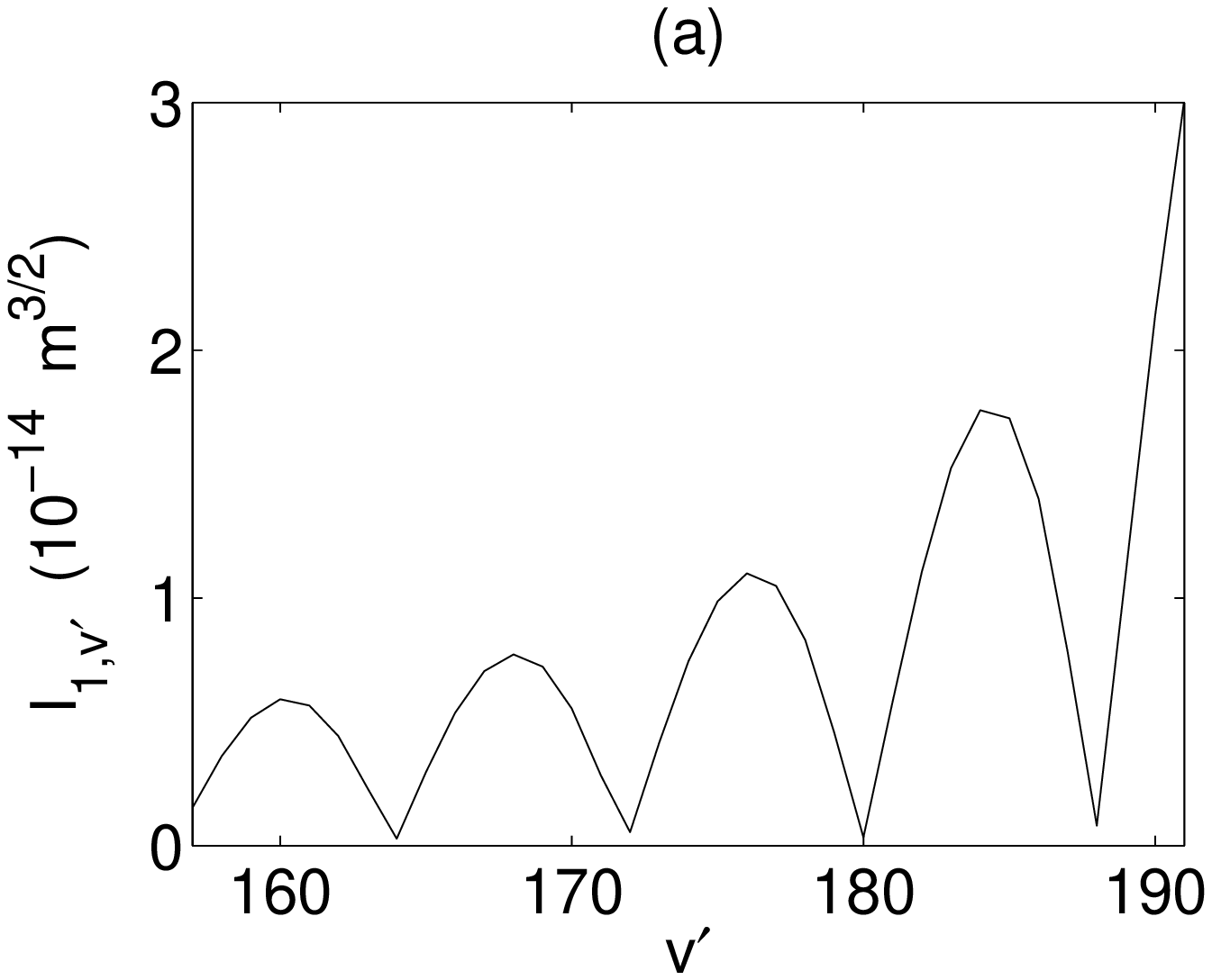} } \par
\par\centering \resizebox*{6cm}{!}{\includegraphics{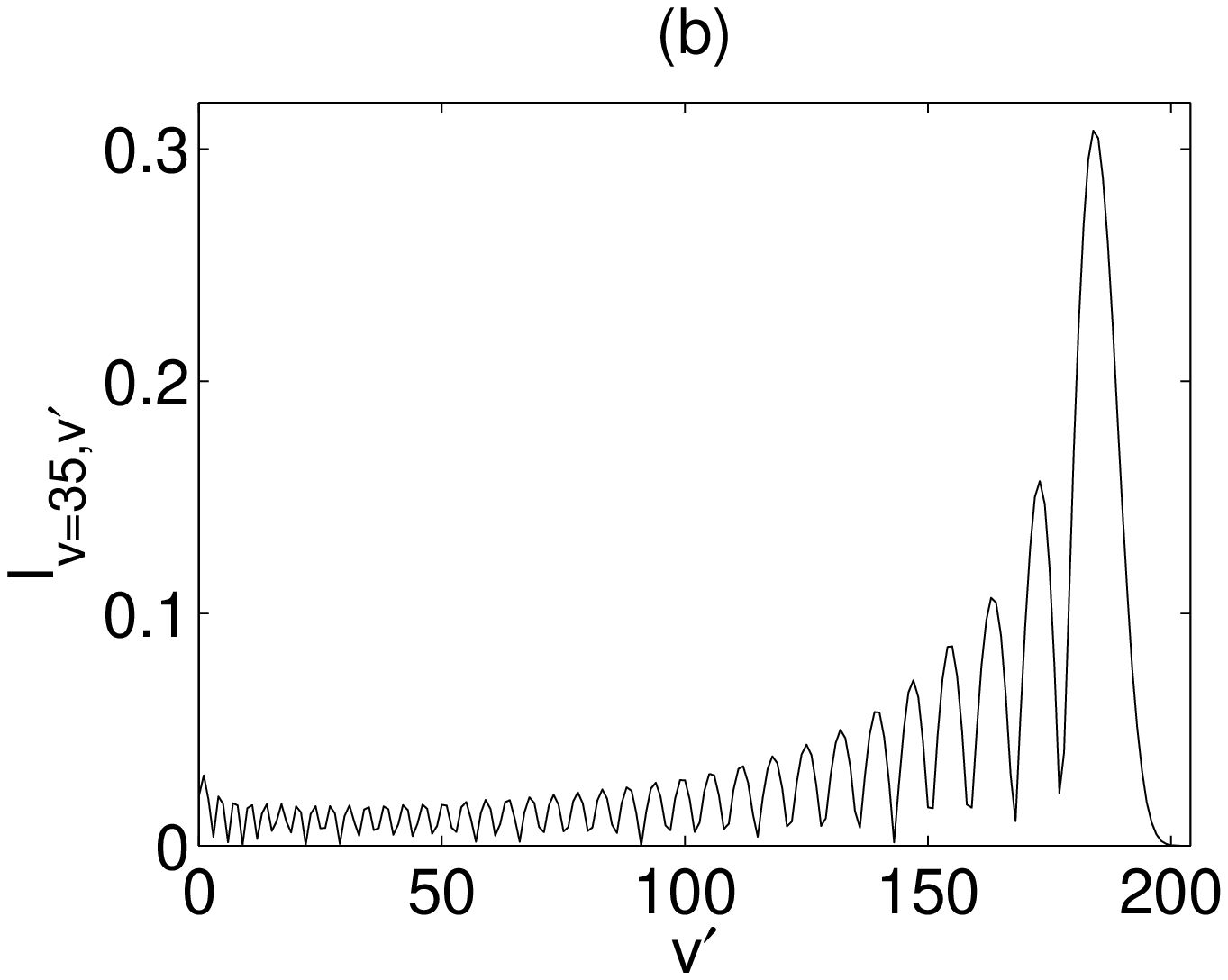} } \par
\caption{Free-bound (a) and bound-bound (b) Franck-Condon overlap integrals,
\protect\( I_{1,v^{\prime }}\protect \) and \protect\( I_{v=35,v^{\prime }}\protect \),
as a function of the vibrational quamtum number \protect\( v^{\prime }\protect \).}
\label{FC overlaps}
\end{figure}

Within such a large range of target levels that the Raman transitions
can be tuned to, several possibilities can be readily eliminated to
simplify the search. For example, Raman transitions via one of the
highly excited levels (\( v^{\prime }\gtrsim 190 \)) will suffer
from large values of the atomic loss coefficient \( \alpha  \) since
the detunings \( D_{i} \) will be small, and as a result the condition
\( \left[ \alpha ^{(i)}\right] ^{-1}\gg T \) will not be satisfied.
On the other hand, transitions via low excited states (\( v^{\prime }\lesssim 160 \))
will have very small values of the free-bound Franck-Condon overlap
integral, \( I_{1,v^{\prime }}\lesssim 0.5\times 10^{-14} \) m\( ^{3/2} \).
This in turn will result in small effective Rabi frequency \( \Omega _{1}^{(eff,0)} \)
(using reasonable values of the intensity of the laser \( \omega _{1} \)
and the density \( n_{1} \)), so that the adiabaticity condition
\( \Omega _{1}^{(eff,0)}T\gg 1 \) is not satisfied.

In general, the behavior of the coefficients \( \Gamma _{1}^{(i)} \),
\( \Gamma _{2}^{(i)} \), and \( \beta _{2}^{(i)} \) is not of a
trivial character. Each particular choice of the final state \( \left| v\right\rangle  \)
in the ground potential which we designate as \( \left| 2\right\rangle  \)
will result in different sets of the detunings of the Raman lasers
with respect to couplings to different excited levels. Further complications
emerge from the oscillatory behavior of the Franck-Condon overlap
integrals, as shown in Fig. \ref{FC overlaps}. This means that the
contribution of the levels nearest to \( \left| 3\right\rangle  \),
having the smallest detunings, may not necessarily give the leading
term in the sums over \( v^{\prime } \), since the further detuned
levels may have larger Franck-Condon overlaps thus resulting in comparable
contributions to the coefficients. In addition, different terms in
the expression for \( \beta _{2}^{(i)} \) will depend on the sign
of the respective detuning, so that they may add up into either a
positive or negative value of \( \beta _{2}^{(i)} \).

Thus our analysis consists of a calculation of all the above coefficients
for different levels \( \left| v\right\rangle  \) and \( \left| v^{\prime }\right\rangle  \),
and subsequent identification of an optimum target level that satisfies
the conditions (\ref{condition-alpha})-(\ref{condition-beta-mismatch-2})
as closely as possible. For each level \( \left| v\right\rangle \equiv \left| 2\right\rangle  \)
in the ground potential we scan the Raman transitions through different
levels \( \left| v^{\prime }\right\rangle  \) in the excited potential,
treating this as \( \left| 3\right\rangle  \) and carrying out the
summations over the remaining levels.

The calculation is done for \( \Omega _{1}^{(el,0)}=10^{10} \) s\( ^{-1} \)
and \( \Omega _{2}^{(el,0)}=10^{9} \) s\( ^{-1} \). To simplify
the analysis we consider the case of \( \Delta _{1}=0 \) and \( \delta =0 \),
i.e. we only treat cases when the primary Raman transition is resonant.
Once the optimum level is identified, we can subsequently fine tune
the two-photon detuning \( \delta  \) for optimum conversion. Provided
that the order of magnitude of the optimum \( \delta  \) is \( |\delta |\sim 10^{4} \)
s\( ^{-1} \) as found in the examples of sub-sections IV. B and C,
this subsequent fine tuning will have no effect on the results of
calculation of the loss and light shift coefficients, since these
involve detunings \( \Delta _{iv^{\prime },\omega _{j}} \) that typically
have much larger magnitudes, \( |\Delta _{iv^{\prime },\omega _{j}}|\gg |\delta | \).

\begin{table}
\begin{tabular}{|l|l|}
\hline 
\( \alpha ^{(1)} \)&
 \( 51.43 \) s\( ^{-1} \)\\
\hline
\( \alpha ^{(2)} \)&
 \( 0.5398 \) s\( ^{-1} \)\\
\hline
\( \Gamma _{1}^{(1)} \)&
 \( 3.010\times 10^{-24} \) m\( ^{3}/ \)s \\
\hline
\( \Gamma _{1}^{(2)} \)&
 \( 3.014\times 10^{-26} \) m\( ^{3}/ \)s \\
\hline
\( \Gamma _{2}^{(1)} \)&
 \( 466.7 \) s\( ^{-1} \)\\
\hline
\( \Gamma _{2}^{(2)} \)&
 \( 4.645 \) s\( ^{-1} \)\\
\hline
\( \beta _{1}^{(1)} \)&
 \( 2.948\times 10^{6} \) s\( ^{-1} \)\\
\hline
\( \beta _{1}^{(2)} \)&
 \( 3.020\times 10^{4} \) s\( ^{-1} \)\\
\hline
\( \beta _{2}^{(1)} \)&
 \( 5.652\times 10^{6} \) s\( ^{-1} \)\\
\hline
\( \beta _{2}^{(2)} \)&
 \( 5.969\times 10^{4} \) s\( ^{-1} \)\\
\hline
\( (\gamma /4)\left| \bar{\Omega }_{2}^{(el,0)}I_{1,3}\right| ^{2}/\left| \Delta _{13,\omega _{2}}\right| ^{2} \)&
 \( 7.96\times 10^{-25} \) m\( ^{3}/ \)s \\
\hline
\( (\gamma /4)\left| \bar{\Omega }_{1}^{(el,0)}I_{2,3}\right| ^{2}/|\Delta _{23,\omega _{1}}|^{2} \)&
 \( 375.9 \) s\( ^{-1} \)\\
\hline
\( |\bar{\Omega }_{1}^{(el,0)}I_{2,3}|^{2}/(4\Delta _{23,\omega _{1}}) \)&
 \( 2.57\times 10^{5} \) s\( ^{-1} \)\\
\hline
\( \chi _{0}'\approx \chi^{\ast} _{0} \)&
 \( -9.25\times 10^{-9} \) m\( ^{3/2} \)/s  \\
\hline
\end{tabular}
\caption{Calculated values of the loss and light shift coefficients for Raman
transitions tuned to the ground \protect\( v=35\protect \) and excited
\protect\( v^{\prime }=177\protect \) levels.}
\label{losses and light shifts}
\end{table}

The most favorable case that we find corresponds to tuning the Raman
transitions to \( v^{\prime }=177 \) in the excited potential bound
by \( -22.23 \) cm\( ^{-1} \) or \( -4.1903\times 10^{12} \) s\( ^{-1} \)
from the dissociation limit, and \( v=35 \) in the ground potential
bound by \( -8.05 \) GHz or \( -5.058\times 10^{10} \) s\( ^{-1} \).
The binding energy of the level \( v=35 \) sets up the frequency
difference \( \omega _{2}-\omega _{1}= \) \( 5.058\times 10^{10} \)
s\( ^{-1} \), and the values of all relevant detunings. The nearby
levels around \( v^{\prime }=177 \) in the excited potential are
separated by about \( 4.2\times 10^{11} \) s\( ^{-1} \), which is
much larger than \( \omega _{2}-\omega _{1} \). For this arrangement
of the target levels, the resonant Franck-Condon overlap integrals
are equal to \( I_{1,3}=1.05\times 10^{-14} \) m\( ^{3/2} \) and
\( I_{2,3}=0.0228 \), so that the effective peak Rabi frequencies
are equal to: \( \Omega _{1}^{(eff,0)}=2.18\times 10^{6} \) s\( ^{-1} \),
for \( n_{1}(0)=4.3\times 10^{20} \) m\( ^{-3} \), and \( \Omega _{2}^{(eff,0)}=2.28\times 10^{7} \)
s\( ^{-1} \). The resulting values of calculated loss and light shift
coefficients are given in Table \ref{losses and light shifts}.

An important factor in minimizing the most significant loss coefficient
\( \Gamma _{2}^{(1)} \) and its {}``cousin'' term in Eq. (\ref{condition-Gamma-2(1)}
) is the relatively small value of \( I_{2,3} \) and of the Franck-Condon
overlaps \( I_{2,v^{\prime }} \) of the closest nearby levels. While
this is favorable for the undesired loss terms, the small value of
\( I_{2,3} \) also affects the strength of the bound-bound coupling
of the primary Raman transition, \( \Omega _{2}^{(eff,0)}=\Omega _{2}^{(el,0)}I_{2,3} \),
which must be kept large. However, the small \( I_{2,3} \)-value
is compensated here by a strong bare electronic Rabi frequency \( \Omega _{2}^{(el,0)}=10^{9} \)
s\( ^{-1} \) so that \( \Omega _{2}^{(eff,0)} \) is still large
and the adiabaticity condition is maintained. 

\begin{figure}
\par\centering \resizebox*{6cm}{!}{\includegraphics{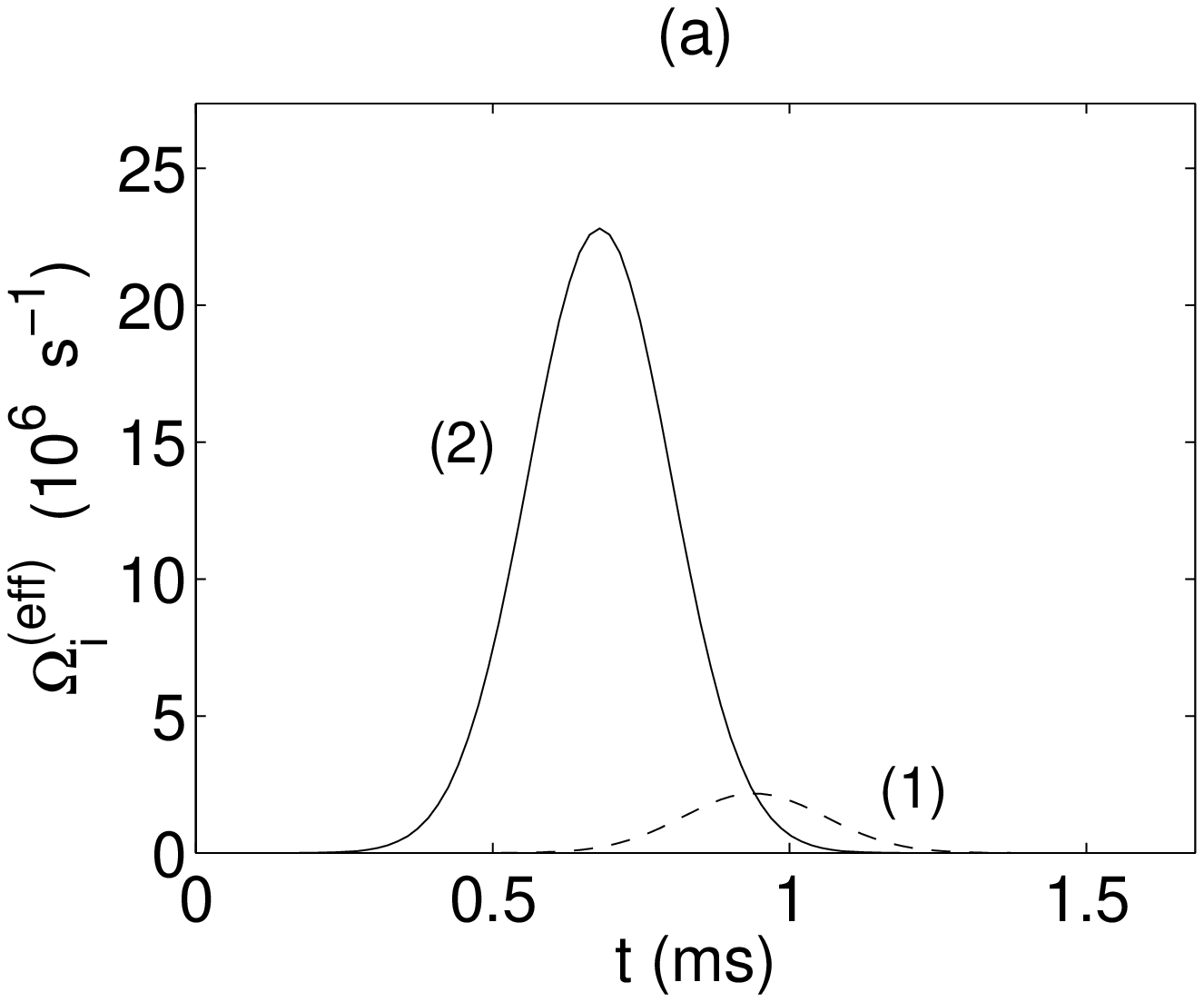} } \par{} \bigskip
\par\centering \resizebox*{6.5cm}{!}{\includegraphics{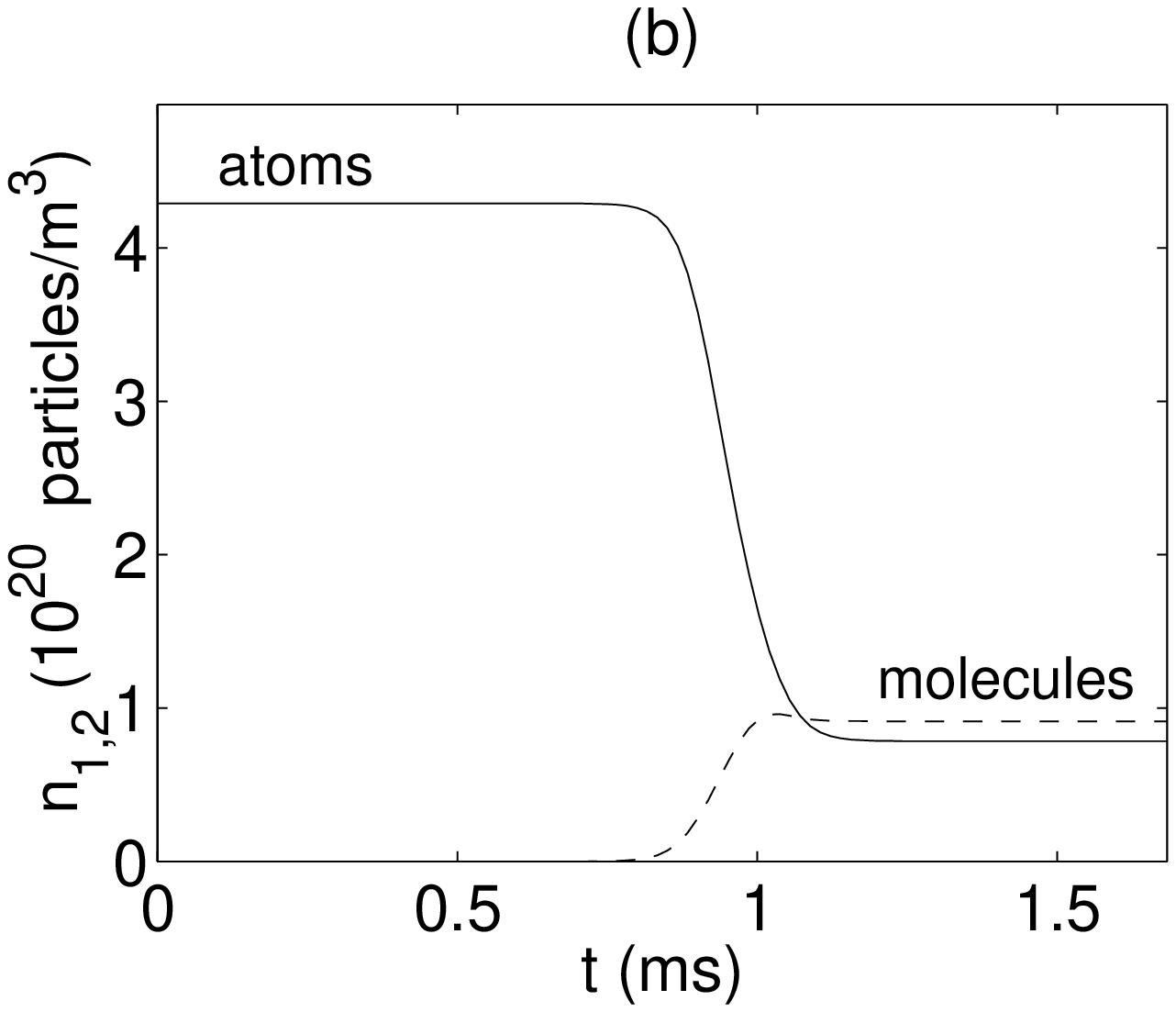} } \par{}
\caption{(a) Optimum effective Rabi frequencies for STIRAP in the multi-level
model. (b) Resulting densities in the uniform condensates case.}
\label{Multi-level}
\end{figure}

Using the above parameter values and simulating the STIRAP equations
with the additional terms given by Eqs. (\ref{losses1})-(\ref{losses3}),
results in about \( 42\% \) conversion efficiency (\( \eta =0.42 \)),
which is rather high and is an encouraging result. In this simulation,
the estimated optimum pulse duration \( T \), the pulse delay coefficient
\( \alpha  \), and the two-photon detuning \( \delta  \), were taken
as follows: \( T=1.7 \) \( \times 10^{-4} \) s\( ^{-1} \), \( \alpha =1.53 \),
and \( \delta =4\times 10^{4} \) s\( ^{-1} \). The effective Rabi
frequencies and the particle number densities for this calculation
are given in Figs. \ref{Multi-level} (a) and (b). We note that these
results are essentially identical whether the exact complex form or
the far-off-resonant approximations to the additional coefficients
are employed. In other words, any coherent cancellation process via
STIRAP in the extra vibrational levels, is negligible compared to
the incoherent losses via non-STIRAP excitation of these levels.

\subsection{Non-uniform condensates}

The final step in our analysis is to include the trap potential and
the kinetic energy terms and simulate the full set of coupled inhomogeneous
mean-field equations (\ref{psi_{3}}) in three space dimensions. The
initial state in these simulations is a pure atomic BEC, with no molecules
present, as given by the standard steady-state Gross-Pitaevskii equation
in a trap. We consider spherically symmetric trap potentials \( V_{i}({\mathbf{x}})=E_{i}+(m_{i}/2)\omega _{i}^{2}|{\mathbf{x}}|^{2} \)
and choose the trap oscillation frequencies \( \omega _{i} \) equal
to each other: \( \omega _{i}/2\pi =100 \) Hz (\( i=1,2,3 \)). Including
these terms, we simulate Eqs. (\ref{psi_{3}}) assuming that the initial
peak density of the atomic BEC is \( n_{1}({\mathbf{x}}=0,t=0)=4.3\times 10^{20} \)
m\( ^{-3} \) at the trap center. This corresponds to the total initial
number of atoms \( N_{1}=\int d^{3}{\mathbf{x}}|\psi ({\mathbf{x}},0)|^{2} \)
equal to \( N_{1}=5\times 10^{5} \).

\begin{figure}
\par\centering \resizebox*{6cm}{!}{\includegraphics{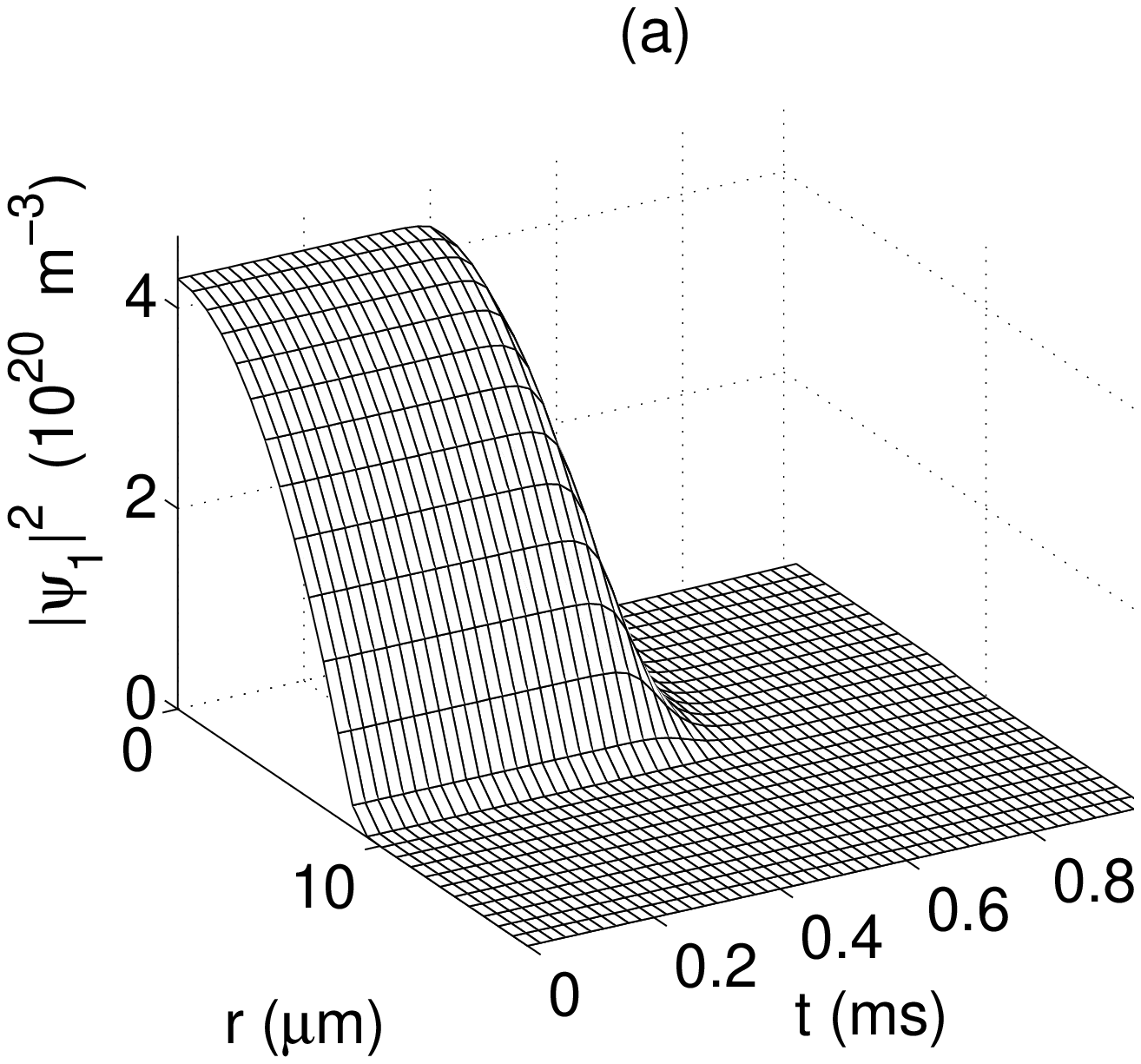} } \par{}
\par\centering \resizebox*{6cm}{!}{\includegraphics{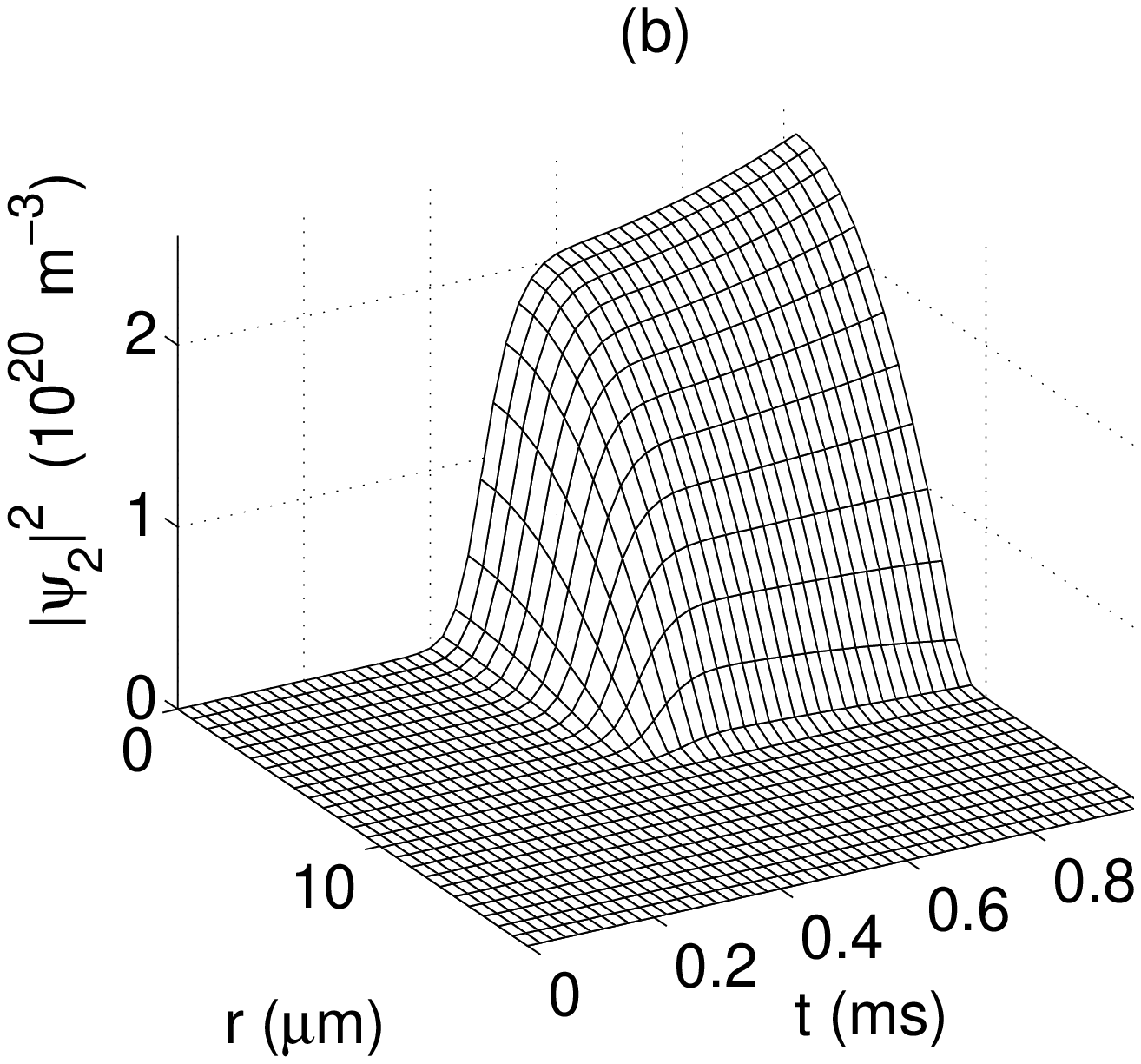} } \par{}
\caption{Densities \protect\( n_{i}({\mathbf{x}},t)=\left| \protect \psi _{i}({\mathbf{x}},t)\right| ^{2}\protect \)
of the atomic (a) and molecular (b) condensates in a trap as a function
of time \protect\( t\protect \) and the radial distance \protect\( r=\left| {\mathbf{x}}\right| \protect \)
from the trap center. The applied pulses are as in Fig. \ref{STIRAP-pulses},
and other parameter values are given in Tables \ref{STIRAP_param}
and \ref{Table_U}. The result is for the idealized three-level model.}
\label{Density}
\end{figure}

We note that the characteristic time scale associated with the trap
potential can be estimated as \( t_{\omega }=1/\omega \simeq 1.6\times 10^{-3} \)
s. The time scale associated with the kinetic energy term is estimated
from the `healing' length \( l_{h}\sim \sqrt{\hbar t_{h}/m_{1}} \)
which corresponds to a healing time scale of \( t_{h} \). Under adiabatic
conversion of an equilibrium (Thomas-Fermi like) atomic BEC, this
healing time scale coincides with the dephasing time \( t_{ph} \)
associated with the mean field energy potential and discussed before.
Both these time scales (\( t_{\omega } \) and \( t_{h} \)) are longer
than the duration of pulses in STIRAP we employed earlier. Therefore,
addition of these terms can not dramatically change the results and
conclusions obtained above for uniform condensates.

\begin{figure}
\par\centering \resizebox*{6cm}{!}{\includegraphics{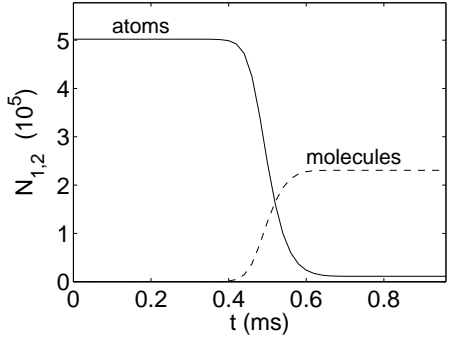} } \par{}
\caption{Integrated occupation numbers \protect\( N_{i}(t)=\int d{\mathbf{x}}\left| \protect \psi _{i}({\mathbf{x}},t)\right| ^{2}\protect \)
of the atomic (solid line) and the molecular (dashed line) fields
as a function of time \protect\( t\protect \), for the parameter
values of Fig. \ref{Density}.}
\label{3D-numbers}
\end{figure}

To show this we first consider the idealized three-level model with
parameter values given in Tables \ref{STIRAP_param} and \ref{Table_U},
i.e. the case of equal and relatively strong effective Rabi frequencies.
The results of simulations of Eqs. (\ref{psi_{3}}) are shown in Figs.
\ref{Density} and \ref{3D-numbers}, where we see a rather high conversion
efficiency \( \eta \simeq 0.91 \). This should be compared with the
uniform condensates result of Fig. \ref{Conversion93}.

Next, we consider the case of more realistic parameter values (lower
Rabi frequencies), and include the effects of incoherent radiative
losses and dephasings as described in the previous sub-section. Using
the calculated values of all relevant parameters for the Raman transitions
tuned to \( v^{\prime }=177 \) and \( v=35 \), and adding the trap
potential and kinetic energy terms to the earlier Eqs. (\ref{losses1})-(\ref{losses3}),
we simulate the process of STIRAP governed by the resulting full set
of mean-field equations in three dimensions: \begin{eqnarray}
\frac{\partial \psi _{1}({\mathbf{x}},t)}{\partial t} & = & i\Delta _{1}^{GP}\psi _{1}+\frac{i{\Omega }^{\ast }_{1}}{\sqrt{2}}\psi _{3}\psi _{1}^{\ast }\nonumber \\
 &  & -\frac{\alpha }{2}\psi _{1}+i\beta _{1}\psi _{1}-\frac{\Gamma _{1}}{2}\left| \psi _{1}\right| ^{2}\psi _{1}+i\bar{U}_{11}\left| \psi _{1}\right| ^{2}\psi _{1}\nonumber \\
 &  & -i\chi \psi _{1}^{\ast }\psi _{2}+i\frac{({\bar{\Omega }}_{2}^{(el)}I_{1,3})^{\ast }}{\sqrt{2}}e^{-i\omega _{12}t}\psi _{1}^{\ast }\psi _{3},
\end{eqnarray}
\begin{eqnarray}
\frac{\partial \psi _{2}({\mathbf{x}},t)}{\partial t} & = & i\Delta _{2}^{GP}\psi _{2}+\frac{i{\Omega }^{\ast }_{2}}{2}\psi _{3}\nonumber \\
 &  & -\frac{\Gamma _{2}}{2}\psi _{2}+i\beta _{2}\psi _{1}\nonumber \\
 &  & -i\frac{\chi ^{\ast }}{2}\psi _{1}^{2}+i\frac{({\bar{\Omega }}_{1}^{(el)}I_{2,3})^{\ast }}{2}e^{i\omega _{12}t}\psi _{3},
\end{eqnarray}
\begin{eqnarray}
\frac{\partial \psi _{3}({\mathbf{x}},t)}{\partial t} & = & i\Delta _{3}^{GP}\psi _{3}+\frac{i{\Omega }_{1}}{2\sqrt{2}}\psi _{1}^{2}+\frac{i{\Omega }_{2}}{2}\psi _{2}\nonumber \\
 &  & +i\frac{{\bar{\Omega }}_{2}^{(el)}I_{1,3}}{2\sqrt{2}}e^{i\omega _{12}t}\psi _{1}^{2}\nonumber \\
 &  & +i\frac{{\bar{\Omega }}_{1}^{(el)}I_{2,3}}{2}e^{-i\omega _{12}t}\psi _{2}.
\end{eqnarray}

We switch on the sequence of two Raman pulses, as shown in Fig. 9,
corresponding to: \( \Omega _{1}^{(eff,0)}=2.18\times 10^{6} \) s\( ^{-1} \),
\( \Omega _{2}^{(eff,0)}=2.28\times 10^{7} \) s\( ^{-1} \), \( T=1.7 \)
\( \times 10^{-4} \) s \( ^{-1} \), and \( \alpha =1.53 \). The
values of \( T \) and \( \alpha  \), together with the choice of
the two-photon detuning \( \delta =4\times 10^{4} \) s\( ^{-1} \), 
correspond to the optimized set of parameters as in Table \ref{Table_main}
and are obtained for \( v^{\prime }=177 \) and \( v=35 \) levels
using \( \Omega _{1}^{(el,0)}=10^{10} \) s\( ^{-1} \) and \( \Omega _{2}^{(el,0)}=10^{9} \)
s\( ^{-1} \). The corresponding values of all other relevant coefficients
are given in Table \ref{losses and light shifts} and \ref{Table_U},
while the spontaneous decay rate is \( \gamma =7.4\times 10^{7} \)
s\( ^{-1} \), as before. The optimum value of \( \delta  \) accounts
for nonzero values of the light shift coefficients \( \beta _{i} \),
so that the actual detuning to be optimized using the Table \ref{Table_main}
is the overall effective detuning \( \tilde{\delta }=\delta +\beta _{2}-2\beta _{1} \).

\begin{figure}
\par\centering \resizebox*{6cm}{!}{\includegraphics{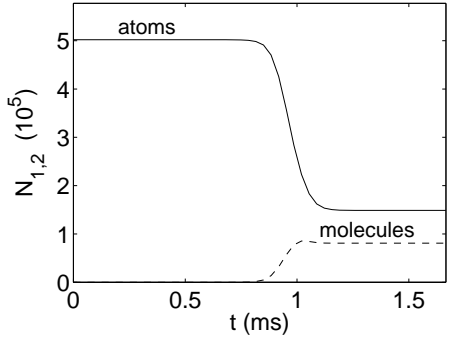} } \par{}
\caption{Same as in Fig. \ref{3D-numbers} but for the parameter values of
the complete multi-level model and a three-dimetional trap geometry.
The resulting conversion efficiency can be compared with the uniform
condensates case of Fig. \ref{Multi-level} (b).}
\label{Final-numbers}
\end{figure}

The results of simulations with this set of parameter values are given
in Fig. \ref{Final-numbers} where we plot the total occupation numbers
in the atomic and molecular BECs as a function of time. The obtained
conversion efficiency is \( \eta \simeq 0.32 \). This is \( 10\% \)
lower than the efficiency in the corresponding homogeneous case of
Fig. \ref{Multi-level}, but still is a rather encouraging result
given the fact that \( 32\% \) conversion of about \( 5\times 10^{5} \)
atoms would give a molecular BEC with the total of \( 8\times 10^{4} \)
molecules, with a peak density of about \( 10^{20} \) m\( ^{-3} \).

\section{Summary}

To summarize, STIRAP is a potential route towards the coherent conversion
of an atomic to a molecular BEC. This process involves stimulated
emission of molecules, and is very different to normal chemical kinetics.
As such, it is a form of `superchemistry' \cite{HWDK2000}. STIRAP
in an atomic BEC can be treated in a very similar way to normal STIRAP,
by introducing an effective Rabi frequency for the first photo-association
transition. A potential difficulty is the relatively low values of
the effective Rabi frequency in the first or photo-association transition.
This can be regarded as physically due to the low densities of atoms
in a typical weakly interacting BEC, compared with atoms in a molecule.
This means that the corresponding Franck-Condon coefficient, after
multiplying by the relevant BEC amplitude, has a very small value.
Thus, the laser intensities required may be quite high, in order to
obtain Rabi frequencies comparable to those used in atomic transitions.

This by itself is not critical, since deep in the adiabatic limit
it is normally permissible to use low Rabi frequencies, as long as
the associated time-scales are long enough. From this point of view,
the use of STIRAP, and consequent reduction of spontaneous emission,
is a physically sensible idea. However, including \( s \)-wave scattering
or mean-field processes into the model for STIRAP sets up a characteristic
two-photon dephasing time scale, so that the pulse durations in STIRAP
have to be \emph{shorter} than a certain critical value. Short pulse
durations necessarily involve high values for the effective Rabi frequencies
in the usual symmetric case of \( \Omega ^{(eff,0)}_{1}=\Omega ^{(eff,0)}_{2} \),
thus requiring a very high laser power for the first transition -
if the adiabaticity condition is to be maintained.

In order to ease this demanding requirement and be able to achieve
highest possible conversion efficiency at smaller total laser power,
we propose to use an off-resonance operation (thus cancelling part
of the mean-field detuning effect) and effective Rabi frequencies
\( \Omega _{1}^{(eff,0)} \) and \( \Omega _{2}^{(eff,0)} \) of different
magnitudes. This has the effect of dramatically increasing coherent
molecule production, in a physically accessible regime of moderate
laser intensity. Further improvements may be possible by tailoring
the input pulse frequencies to the time-dependent two-photon detuning,
caused by inter-atomic and intermolecular scattering. We have carried
out mean-field calculations in three dimenisons to verify that trap
inhomogeneity should not have adverse effects on the STIRAP process.

Finally, we stress the importance of radiative losses and dephasing
due to incoherent processes that occur during STIRAP. These processes are
usually assumed to be negligible in ordinary STIRAP between purely
atomic or molecular states. In the present case of coupled atomic/molecular
BECs, this assumption can not be easily justified since the free-bound
transition typically involves a relatively low effective Rabi frequency,
which necessitates long pulse durations. Instead, we find that the
incoherent couplings can be rather destructive unless special care
is taken to minimize their effect. This involves detailed knowledge
of the structure of the free-bound and bound-bound transitions and
subsequent identification of optimum target levels in STIRAP, so that
the overall conversion efficiency remains comparable to the predictions
of the simplified three-level model.

\section*{Acknowledgments}

P. D. and K. K. gratefully acknowledge the ARC for the support of
this work. D. H. and R. W. acknowledge the support by the U. S. National
Science Foundation, The R. A. Welch Foundation, and the NASA Microgravity
Research Program.

\end{document}